\documentclass[graybox]{svmult}

\usepackage{type1cm}        % activate if the above 3 fonts are
\usepackage[utf8]{inputenc}
\usepackage{textgreek}
\usepackage{makeidx}         % allows index generation
\usepackage{graphicx}        % standard LaTeX graphics tool
\usepackage{multicol}        % used for the two-column index
\usepackage[bottom]{footmisc}% places footnotes at page bottom

\usepackage{newtxtext}       % 
\usepackage{newtxmath}       % selects Times Roman as basic font
\usepackage{multirow}
\usepackage{lscape}                                                                           
\usepackage{soul}

\makeindex             % used for the subject index
\newcommand{\red}[1]{\textcolor{red}{\bf #1}}    % SC
\begin{document}

\title*{Statistical characterization and classification of astronomical transients with Machine Learning in the era of the Vera C. Rubin Observatory}
\titlerunning{Machine learning based classification of transients in the era of LSST}
\author{Marco Vicedomini, Massimo Brescia, Stefano Cavuoti, Giuseppe Riccio, Giuseppe Longo}
\authorrunning{Vicedomini, Brescia, Cavuoti et al. 2020}
\institute{M. Vicedomini, S. Cavuoti and G. Longo \at Department of Physics, University of Naples  Federico II, Strada Vicinale Cupa Cintia, 21, I-80126 Napoli, Italy. \email{stefano.cavuoti@inaf.it} \and
 M. Brescia and G. Riccio \at INAF - Astronomical Observatory of Capodimonte, Salita Moiariello 16, I-80131 Napoli, Italy. \email{massimo.brescia@inaf.it}}
\maketitle

\abstract*{Astronomy has entered the multi-messenger data era and Machine Learning has found widespread use in a large variety of applications. The exploitation of synoptic (multi-band and multi-epoch) surveys, like LSST (Legacy Survey of Space and Time), requires an extensive use of automatic methods for data processing and interpretation. With data volumes in the petabyte domain, the discrimination of time-critical information has already exceeded the capabilities of human operators and crowds of scientists have extreme difficulty to manage such amounts of data in multi-dimensional domains. 
This work is focused on an analysis of critical aspects related to the approach, based on Machine Learning, to variable sky sources classification, with special care to the various types of Supernovae, one of the most important subjects of Time Domain Astronomy, due to their crucial role in Cosmology. The work is based on a test campaign performed on simulated data. The classification was carried out by comparing the performances among several Machine Learning algorithms on statistical parameters extracted from the light curves. The results make in evidence some critical aspects related to the data quality and their parameter space characterization, propaedeutic to the preparation of processing machinery for the real data exploitation in the incoming decade.}

\textit{\footnotesize{Preprint version of the manuscript to appear in the Volume “Intelligent Astrophysics”of the series “Emergence, Complexity and Computation”, Book eds. I. Zelinka, D.Baron, M. Brescia, Springer Nature Switzerland, ISSN: 2194-7287}}\\

\abstract{Astronomy has entered the multi-messenger data era and Machine Learning has found widespread use in a large variety of applications. The exploitation of synoptic (multi-band and multi-epoch) surveys, like LSST (Legacy Survey of Space and Time), requires an extensive use of automatic methods for data processing and interpretation. With data volumes in the petabyte domain, the discrimination of time-critical information has already exceeded the capabilities of human operators and crowds of scientists have extreme difficulty to manage such amounts of data in multi-dimensional domains. 
This work is focused on an analysis of critical aspects related to the approach, based on Machine Learning, to variable sky sources classification, with special care to the various types of Supernovae, one of the most important subjects of Time Domain Astronomy, due to their crucial role in Cosmology. The work is based on a test campaign performed on simulated data. The classification was carried out by comparing the performances among several Machine Learning algorithms on statistical parameters extracted from the light curves. The results make in evidence some critical aspects related to the data quality and their parameter space characterization, propaedeutic to the preparation of processing machinery for the real data exploitation in the incoming decade.}

%%%%%%%%%%%%%%%%%%%%%%%%
\section{Introduction}
\label{sec:Intro}

The scientific topics covered in this work falls within what is called Time Domain Astronomy. This is the study of variable sources, i.e. astronomical objects whose light changes with time. Although the taxonomy of such sources is extremely rich, there are two main kinds of objects, respectively, transients and variables. The first changes its nature during the event, while the second presents just a brightness variation. The study of these phenomena is fundamental to identify and analyze either the mechanisms causing light variations and the progenitors of the various classes of objects.

Since ancient times the phenomenon of Supernovae (SNe) has fascinated human beings, but only recently we understood, in most cases, why and how this explosion happens \cite{Branch2010}. Obviously there are still many open questions, but the knowledge about the type of galaxy hosting various kinds of Supernova and at which rate they take place, could help us to better understand this phenomenon and many other related properties of the Universe \cite{Goobar2011}. 

For example, the observed luminosity dispersion of SNe is evidenced through inhomogeneities in the weak lensing event and this is an upper limit on the cosmic matter power spectrum. Massive cosmological objects like galaxies and clusters of galaxies can magnify many times the flux of events like SNe that would be too faint to detect and bring them into our analysis scope. Studies on lensed SNe type Ia by clusters of galaxies may be used to probe the distribution of dark matter on them. Time delay between the multiple images of lensed SNe could provide a good estimates of its high redshift. Furthermore there are two factors that makes SNe better than other sources, like quasars, in measuring time delay \cite{Huber2019}: \textit{(i)} if the Supernovae is taken before the peak, the measurements are easier and on short timescale compared to the quasars; \textit{(ii)} the SN light fade away with time, so we can measure the lens stellar kinematics and the dynamics lens mass modeling. In the next decade, the Vera C. Rubin Observatory will perform the Rubin Observatory Legacy Survey of Space and Time (LSST), using the Rubin Observatory LSST Camera and the Simonyi Survey Telescope. LSST  will play a key role in the discovery of new lensed SNe Ia  \cite{Ivezic2011}. LSST will help to find apparently host-less SNe of every type, and this may help to study dwarf galaxies with a mass range of $10^4 \div 10^6$ solar masses. These galaxies, indeed, play a key role in large scale structure models, and despite their very big predicted population, over $1$ Mpc we cannot see them until now. Same story for the theorized intracluster population of stars stripped from their galaxies, which could be seen through the SNe host-less events.

%%%%%%%%%%%%%%%%%%%%%%%%%%%%%%%%%%%
In order to understand and push ourselves further and further into the universe, ever more powerful incoming observing instruments, like LSST, will be able to deliver impressive amounts of data, for which astronomers are obliged to make an intensive use of automatic analysis systems. Methods that fall under the heading Data Mining and Machine Learning have now become commonplace and indispensable to the work of scientists \cite{Brescia2018, Brescia2017, Brescia2013}. But then, where the human work is still needed? For sure in terms of final analysis and validation of the results. This thesis work is therefore based on this virtuous combination, by exploiting data science methodology and models, such as Random Forest \cite{Breiman2003}, Nadam, RMSProp and Adadelta \cite{Dozat2015}, to perform a deep investigation on time domain astronomy, by focusing the attention on Supernovae classification, performed on realistic sky simulations. Furthermore, a special care has been devoted to the parameter space analysis, through the application of the method $\Phi$LAB \cite{Brescia2019,DV2019} to the various classification experiments, in order to evaluate the commonalities among them in terms of features found as relevant to solve the recognition of different types of transients.

%%%%%%%%%%%%%%%%%%%%%%%%%%%%%%%%%%%%%%%%%%
In Sec.~\ref{sec:Data} we describe the two data simulations used for the experiments and the extracted statistical features composing the data parameter space. In Sec.~\ref{sec:Models} we give a brief introduction of the ML methods used, while in Sec.~\ref{sec:Experiments} the series of experiments performed are deeply reported. Finally, in Sec.~\ref{discussion} we analyze the results and draw the conclusions.

%%%%%%%%%%%%%%%%%%%%%%%%
\section{Data}
\label{sec:Data}
In this work two simulation datasets were used; the Supernova Photometric Classification Challenge (hereafter SNPhotCC, \cite{Kessler2010}) and the Photometric LSST Astronomical Time-Series Classification Challenge (hereafter PLAsTiCC, \cite{PLAS2020,PLAS12019,plasticc}). 

%%%%%%%%%%%%%%%%%%%%%%%%
\subsection{The SNPhotCC simulated catalogue}
\label{sec:Data-SNPhotCC}
This catalogue was the subject of a challenge performed in 2010 and consists of a mixed set of simulated SN types, respectively, Ia, Ibc and II, selected by respecting the relative rate (Table~\ref{SNPtable}). The volumetric rate was found by Dilday et al. \cite{Dilday2008} as $r_v=\alpha(1+z)^{\beta}$, where for SNe Ia parameters we have $\alpha_{Ia}=2.6\times10^{-5} Mpc^{-3}h_{70}^3$ $yr^{-1}$, $\beta_{Ia}=1.5$ and $h_{70}= H_0/(70~ km s^{-1} Mpc^{-1})$. $H_0$ is the present value of the Hubble parameter. For non Ia SNe, the parameters come from Bazin et al. \cite{Bazin2009} and are $\alpha_{NonIa}=6.8\times10^{-5} Mpc^{-3}h_{70}^3$ $yr^{-1}$ and $\beta_{NonIa}=3.6$. The simulation is based on four bands, \textit{griz}, with cosmological parameters $\Omega_{M}=0.3$, $\Omega_{\Lambda}=0.7$ and $\omega=-1$, where $\Omega_{M}$ is the density of barionic and dark matter, $\Omega_{\Lambda}$ is the density of dark energy and $\omega$ is the cosmological constant. Moreover, the point-spread function, atmospheric transparency and sky-noise were measured in each filter and epoch using the one-year chronology. 

\begin{figure}
\includegraphics[width=\textwidth]{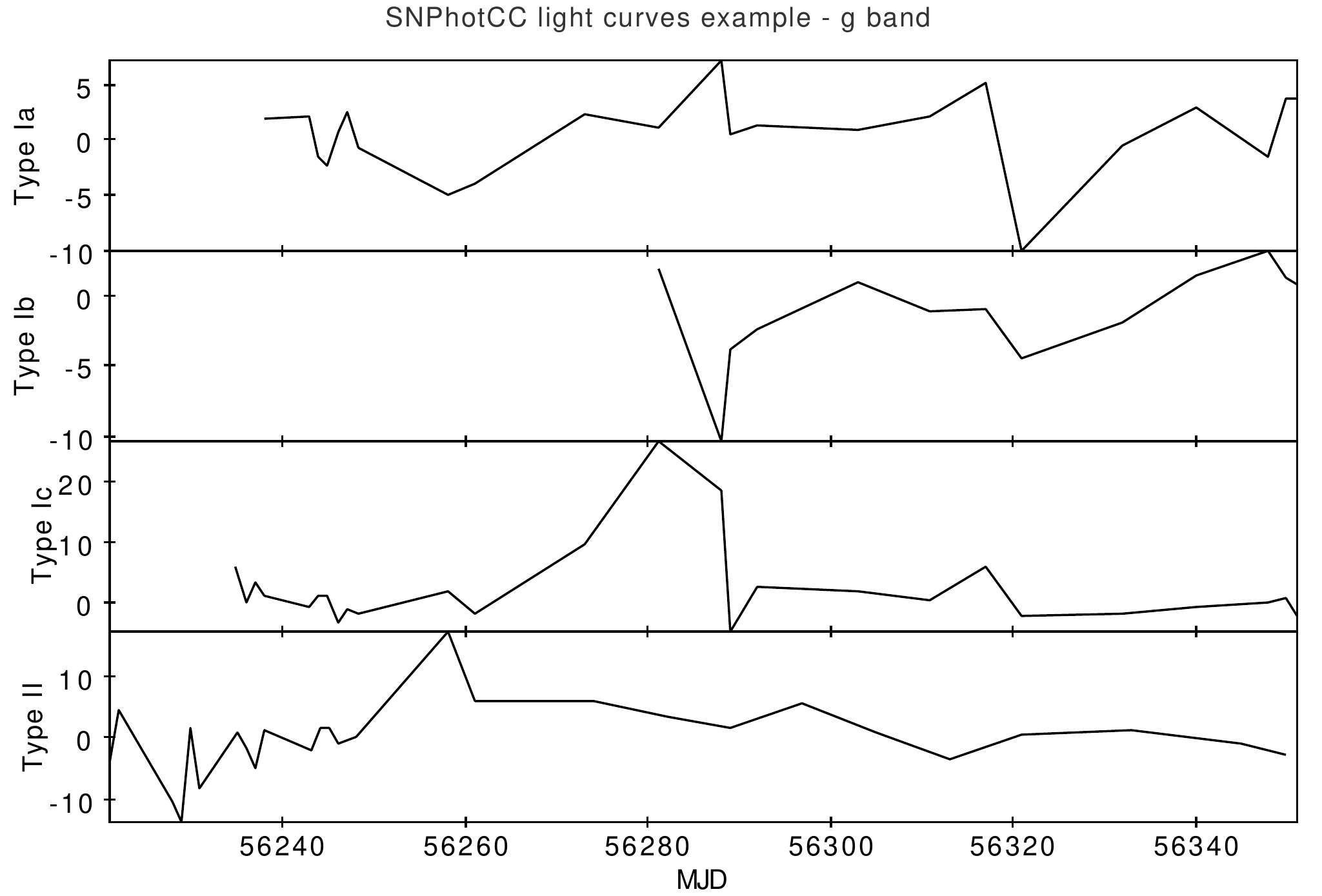}
\caption{Examples of SNPhotCC light curves in \textit{g} band. From the top to the bottom: SN004923(Ia), SN000760(Ib), SN003475(Ic), SN001986(II).}
\label{SNPfigure}
\end{figure}

\begin{table}
  \centering
 \begin{tabular}{||c c c c c||} 
 \hline
 Types & Bands & Sampling & \% & Amount \\ [0.5ex] 
 \hline\hline
 SNIa & g,r,i,z & uneven & 23,86 & 5088 \\ 
 \hline
 SNIbc & g,r,i,z & uneven & 13,14 & 2801 \\
 \hline
 SNII & g,r,i,z & uneven & 63 & 13430 \\
 \hline
\end{tabular}
\caption{SNPhotCC dataset composition.}
\label{SNPtable}
\end{table}

The dataset sources are based on two variants, respectively, with or without the host-galaxy photometric redshift. For this work only the samples without redshift information were used. 

Every simulated light curve has at least one observation, in two or more bands, with signal-to-noise ratio $>5$ and five observations after the explosion (Fig.~\ref{SNPfigure}). A spectroscopically confirmed training subset was provided; it was based on observations from a 4m class telescope with a limiting r-band magnitude of 21.5 and on observations from an 8m class telescope with a limiting \textit{i}-band magnitude of 23.5.

%%%%%%%%%%%%%%%%%%%%%%%%
\subsection{The PLAsTiCC simulated catalogue}
\label{sec:Data-PLAsTiCC}

This catalogue arises from a challenge focused on the future use of the LSST\footnote{\url{https://www.kaggle.com/c/PLAsTiCC-2018}}, by simulating the possible objects on which science will be based. In particular, most of these objects are transients.

LSST will be the largest telescope specialized for the Time Domain Astronomy, whose first light is foreseen in late 2020. Its field of view will be $\sim3.5$ degrees (the diameter will be about seven full moons side by side), with a $6.5$m effective aperture, a focal ratio of $1.23$ and a camera of $3.2$ Gigapixel.

Every four nights it will observe the whole sky visible from the Chile (southern emisphere). Therefore, it will find an unprecented amount of new transients: Supernovae Ia, Ia-91bg, Iax, II, Ibc, SuperLuminous (SL), Tidal Disruption Events, Kilonova, Active Galactic Nuclei, RR Lyrae, M-dwarf stellar flares, Eclipsing Binary and Pulsating variable stars, $\mu$-lens from single lenses, $\mu$-lens from binary lenses, Intermediate Luminosity Optical Transients, Calcium Rich Transients and Pair Instability Supernovae.

LSST data will be used for studying stars in our Galaxy, understanding how solar systems and galaxies formed and the role played by massive stars in galaxy chemistry as well as measuring the amount of matter in the Universe. PLAsTiCC includes light curves with realistic time-sampling \cite{plasticc}, noise properties and realistic astrophysical sources.

\begin{figure}
\includegraphics[width=\textwidth]{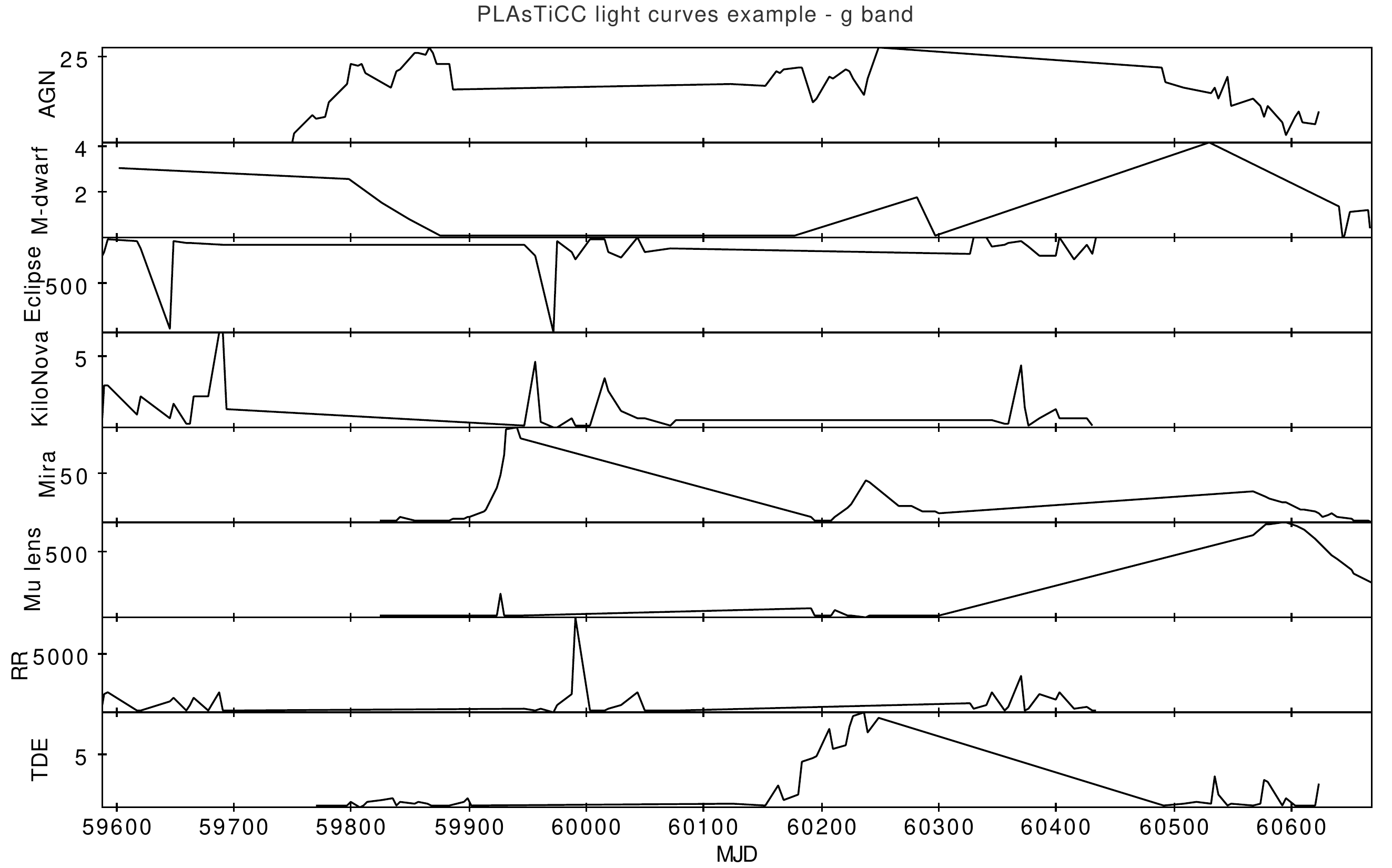}
\caption{Examples of PLAsTiCC light curves in \textit{g} band. From the top to the bottom: 2198(AGN), 2157270(M-Dwarf), 22574(Eclipsing Binary), 139362(Kilonova), 80421(Mirae), 45395($\mu$-lens), 184176(RR lyrae), 9197(TDE).}
\label{PLAfigure1}

\medskip

\includegraphics[width=\textwidth]{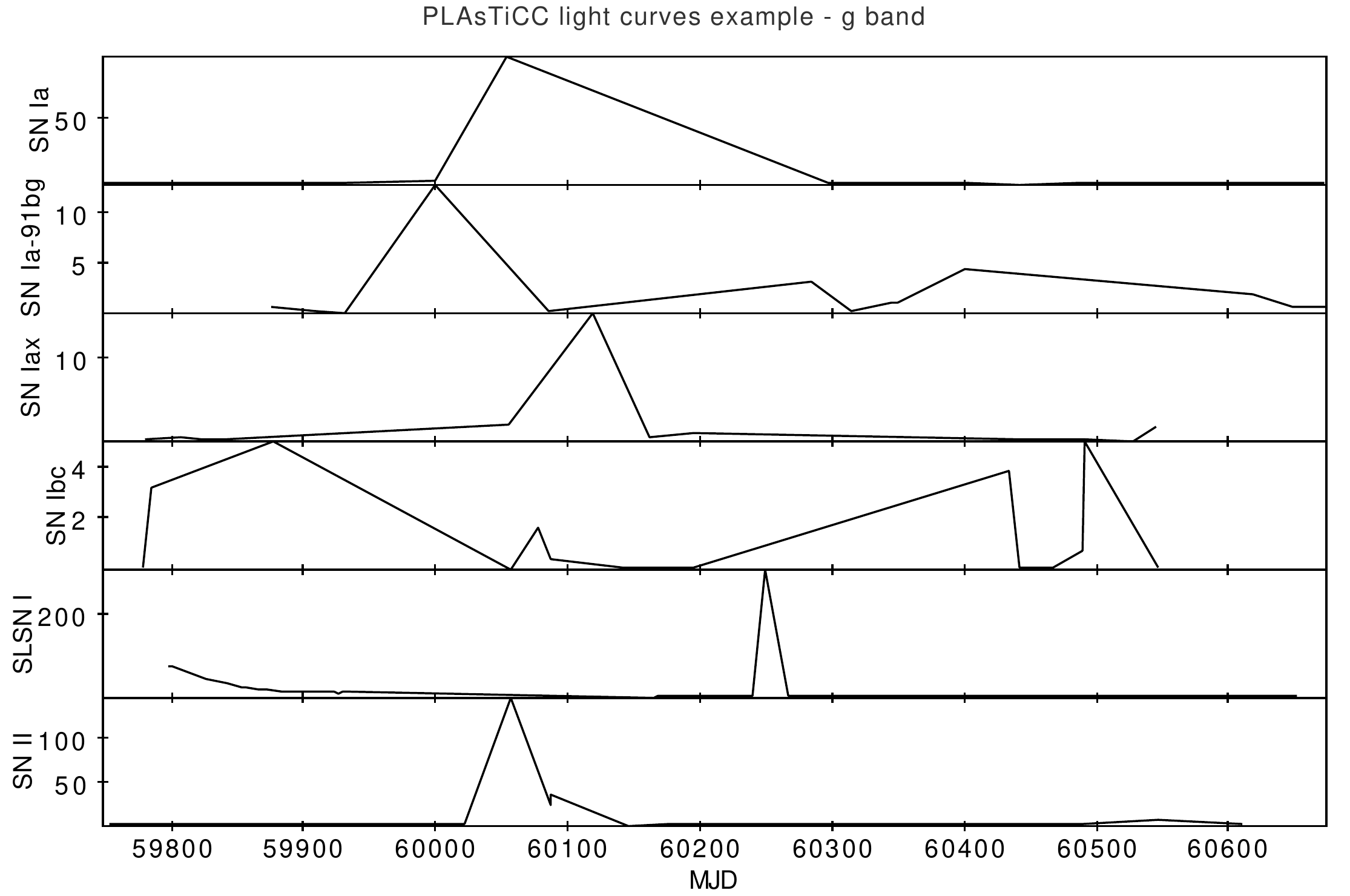}
\caption{Examples of PLAsTiCC light curves in \textit{g} band. From the top to the bottom: 15461391(SNIa), 1143209(SNIa-91bglike), 1019556(SNIax), 1076072(SNIbc), 73610(SLSN I), 1028853(SNII).}
\label{PLAfigure2}
\end{figure}

Each object has observations in six bands: \textit{u} ($300\div400 $ nm), \textit{g} ($400\div600 $ nm), \textit{r} ($500\div700 $ nm), \textit{i} ($650\div850 $ nm), \textit{z} ($800\div950 $ nm), and \textit{y} ($950\div1050 $ nm). 
The training set is a mixture of what we can expect to have before LSST, so it is a quite homogeneous ensemble of $\sim8000$ objects; the test set, instead, is based on what we expect to have after 3 years of LSST operations and it is formed by $\sim 3,5$ million of objects. The observations are limited in magnitude in single band to 24.5 in the \textit{r} band and to 27.8 \textit{r} stacked band (see Figures \ref{PLAfigure1} and \ref{PLAfigure2} for examples of light curves). By combining training and test, we collected the objects per class as listed in Table~\ref{PLAtable}.

\begin{table}
  \centering
 \begin{tabular}{||c c c c c c c||} 
 \hline
 Types & Training & Test & Bands & Sampling & \% & Amount \\ [0.5ex] 
 \hline\hline
 SNIa & 2313 & 1659831 & u,g,r,i,z,y & uneven & 47.57 & 1662144\\ 
 \hline
 SNIax & 183 & 63664 & u,g,r,i,z,y & uneven & 1.81 & 63847\\
 \hline
 SNIa 91bglike & 208 & 40193 & u,g,r,i,z,y & uneven & 1.15 & 40401\\
 \hline
 SNIbc & 484 & 175094 & u,g,r,i,z,y & uneven & 5.00 & 175578\\ 
 \hline
 SNII & 1193 & 1000150 & u,g,r,i,z,y & uneven & 28.65 & 1001343\\ 
 \hline
 SLSN I & 175 & 35782 & u,g,r,i,z,y & uneven & 1.02 & 35957\\ 
 \hline
 AGN & 370 & 101424 & u,g,r,i,z,y & uneven & 2.89 & 101794\\ 
 \hline
 M-Dwarf & 981 & 93494 & u,g,r,i,z,y & uneven & 2.68 & 94475\\ 
 \hline
 RR Lyrae & 239 & 197155 & u,g,r,i,z,y & uneven & 5.63 & 197394\\ 
 \hline
 Mirae & 30 & 1453 & u,g,r,i,z,y & uneven & 0.04 & 1483\\ 
 \hline
 Eclipse & 924 & 96572 & u,g,r,i,z,y & uneven & 2.77 & 97496\\ 
 \hline
 KN & 100 & 131 & u,g,r,i,z,y & uneven & 0.01 & 231\\ 
 \hline
 TDE & 495 & 13555 & u,g,r,i,z,y & uneven & 0.38 & 14050\\ 
 \hline
 $\mu$ Lens & 151 & 1303 & u,g,r,i,z,y & uneven & 0.04 & 1454\\ 
 \hline
 Other & 0 & 13087 & u,g,r,i,z,y & uneven & 0.36 & 13087\\ 
 \hline
\end{tabular}
\caption{PLAsTiCC dataset composition.}
\label{PLAtable}
\end{table}

%%%%%%%%%%%%%%%%%%%%%%%%
\subsection{The statistical parameter space}
\label{sec:Data-features}
In order to evaluate the classification performances, the light curves of the objects have been subject of a statistical approach, by transforming them into a set of features representing some peculiar characteristics of the astrophysical objects. Within this work we used the following features (already used in a similar task in \cite{DIsanto2016}), resulting from a preliminary mapping of variable object light curves into a statistical parameter space:

\begin{itemize}
\item Amplitude (ampl): the arithmetic average between the maximum and the minimum magnitude,
\begin{equation}
ampl=\frac{mag_{max}-mag_{min}}{2} 
\end{equation}

\item Beyond1std (b1std): the fraction of photometric points above or under one standard deviation from the weighted average,
\begin{equation}
b1std= P(|mag-\overline{mag}|> \sigma)
\end{equation}

\item Flux Percentage Ratio (fpr): the ratio between two flux percentiles $F_{n,m}$. The flux percentile is defined as the difference between the flux value at percentiles n and m, respectively. For this work, the following fpr values have been used:
\begin{gather*}
	fpr20=F_{40,60}/F_{5,95} \\
	fpr35=F_{32,5,67,5}/F_{5,95}\\
	fpr50=F_{25,75}/F_{5,95}\\
	fpr65=F_{17,5,82,5}/F_{5,95}\\
	fpr80=F_{10,90}/F_{5,95}
\end{gather*}

\item Lomb-Scargle Periodogram (ls): the period obtained by the peak frequency of the Lomb-Scargle periodogram.\\

\item Linear Trend (lt): the slope \textit{a} of the light curve in the linear fit,
\begin{equation*}
mag= a*t+b
\end{equation*}
\begin{equation}
lt= a
\end{equation}

\item Median Absolute Deviation (mad): the median of the deviation of fluxes from the median flux,
\begin{equation}
mad= median_i(|x_i-median_j(x_j)|)
\end{equation}

\item Median Buffer Range Percentage (mbrp): the fraction of data points which are within 10\% of the median flux,
\begin{equation}
mbrp= P(|x_i-median_j(x_j)|<0.1*median_j(x_j))
\end{equation}

\item Magnitude Ratio (mr): an index to see if the majority of data points are above or below the median of the magnitudes,
\begin{equation}
mr= P(mag>median(mag))
\end{equation}

\item Maximum Slope (ms): the maximum difference obtained measuring magnitudes at successive epochs,
\begin{equation}
ms= max(|\frac{(mag_{i+1}-mag_i)}{(t_{i+1}-t_i)}|)=\frac{\Delta mag}{\Delta t}
\end{equation}

\item Percent Difference Flux Percentile (pdfp): the difference between the fifth and the 95th percentile flux, converted in magnitudes, divided by the median flux,
\begin{equation}
pdfp= \frac{(mag_{95}-mag_5)}{median(mag)}
\end{equation}

\item Pair Slope Trend (pst): the percentage of the last 30 couples of consecutive measures of fluxes that show a positive slope,
\begin{equation}
pst= P(x_{i+1}-x_i>0,i=n-30,...,n)
\end{equation}

\item R Cor Bor (rcb): the fraction of magnitudes that is above 1.5 magnitudes with respect to the median,
\begin{equation}
rcb= P(mag>(median(mag)+1.5))
\end{equation}

\item Small Kurtosis (kurt): the ratio between the 4th order momentum and the square of the variance. For small kurtosis it is intended the kurtosis on a small number of epochs,
\begin{equation}
kurt= \frac{\mu_4}{\sigma^2}
\end{equation}

\item Skewness (skew): the ratio between the 3rd order momentum and the variance to the third power,
\begin{equation}
skew= \frac{\mu_3}{\sigma^3}
\end{equation}

\item Standard deviation (std): the standard deviation of the flux.
\end{itemize}

%%%%%%%%%%%%%%%%%%%%%%%%
\section{Machine Learning models}
\label{sec:Models}
A classifier can be used as a descriptive model to distinguish among objects of different classes, and as a predictive model to predict the class label of input patterns. Classification techniques work better for predicting or describing data sets with binary or nominal categories. 
Each technique uses a different learning algorithm to find a model that fits the relationship between the feature set and class labels of the input data. The goal of the learning algorithm is to build models with good generalization capability.
The typical approach of machine learning models is to randomly shuffle and split the given input dataset with known assigned class labels into three subsets: training, validation and blind test sets. The validation set can be used to validate the learning process, while the test set is used blindly to verify the trained model performance and generalization capabilities.
In the following sections we briefly introduce the methods used to perform the classification experiments, together with the statistical estimators adopted to evaluate their performances.

%%%%%%%%%%%%%%%%%%%%%%%%
\subsection{The Random Forest classifier}
\label{sec:RF}
A Random Forest (RF, \cite{Breiman2003}) is a classifier consisting of a collection of tree-structured classifiers $\{h(x,\Theta_k), k=1, ...\}$ where the $\{\Theta_k \}$ are independent identically distributed random vectors and each tree casts a unit vote for the most popular class at input x \cite{Breiman2001}.
The generalization error for this algorithm depends on the strength of single trees and from their correlations through the raw margin functions. The upper bound, instead, tell us that smaller the ratio of those quantities, better the RF performance. To improve the model accuracy by keeping trees strength, the correlation between trees is decreased and bagging with a random selection of features is adopted. Bagging or Bootstrap Aggregating, is a method designed to improve the stability and accuracy of machine learning algorithms. It also reduce variance and minimizes the risk of overfitting.
Given a training set of size n, bagging generates m new training sets, each of size p, by sampling from the original one uniformly and with replacement. This kind of sampling is known as a bootstrap sample. The m models are fitted using the m bootstrap samples and combined by averaging the output (for regression) or voting (for classification). Bagging is useful because, in addition to improving accuracy when using random features, it provides an estimate of the generalized error of the set of trees and the strength and correlation of trees. The estimation is done out-of-bag. Out-of-bag means that the error estimate of each pair (x,y) is made on all those bagging datasets that do not contain that given pair.

%%%%%%%%%%%%%%%%%%%%%%%%
\subsection{The Nadam, RMSProp and Adadelta classifiers}
\label{sec:NADAM}
The simplest optimization algorithm is the \textit{Gradient Descent}, in which the gradient of the function to be minimized is calculated. This depends on the parameter $\theta_{t-1}$. Only a portion of the gradient is used to update the parameters; this portion is given by the parameter $\eta$:

\begin{equation*}
\begin{cases}
g_t \longleftarrow \nabla_{\theta_{t-1}}f(\theta_{t-1}-\eta \mu m_{t-1})\\
m_t \longleftarrow \mu m_{t-1}+g_t\\
\theta_t \longleftarrow \theta_{t-1}-\eta m_t
\end{cases}
\end{equation*}

where \textit{m} is the so-called \textit{momentum vector}, used to accelerate the update of the learning function, while $\mu$ is the decay constant. These two terms increase the speed of gradient decreasing in the direction where the gradient tends to remain constant, while reducing it where the gradient tends to oscillate.

Nadam is a modified version of the Adam algorithm, based on the combination between the momentum implementation and the $L_2$ normalization. This type of normalization changes the $\eta$ member, dividing it by the $L_2$ norm of all previous gradients.

Adadelta is a variant that tries to reduce the aggressive, monotonically decreasing learning rate. In fact, instead of accumulating all past squared gradients, it restricts the window of accumulated past gradients to some fixed size w.
This has the advantage of compensating for the speeds along the different dimensions by stabilizing the model on common features and allowing the rare ones to emerge. A problem of this algorithm comes from the norm vector that could become so large to stop the training, preventing the model from reaching the local minimum. This problem is solved by RMSProp, a $L_2$ normalization based algorithm, which replaces the sum of $n_t$ with a decaying mean, characterized by a costant value $\nu$. This allows the model to avoid any stop of the learning process. For a detailed description of these models, see \cite{Dozat2015}.
 
%%%%%%%%%%%%%%%%%%%%%%%%
\subsection{Parameter Space exploration}
\label{sec:philab}
The choice of an optimal set of features is connected to the concept of feature importance, based on the measure of a feature’s relevance \cite{DV2019}. Formally, the importance or relevance of a feature is its percentage of informative contribution to a learning system.
We approached the feature selection task in terms of the \textit{all-relevant} feature selection, able to extract the most complete parameter space, i.e. all features considered relevant for the solution to the problem. This is appropriate for problems with highly correlated features, as these features will contain nearly the same information. With a minimal-optimal feature selection, choosing any one of them (which could
happen at random if they are perfectly correlated), means that the rest will never be selected.
The method $\Phi$LAB, deeply discussed in \cite{DV2019}, includes properties of both embedded and wrappers categories of feature selection to optimize the parameter space, by solving the \textit{all-relevant} feature selection problem, thus indirectly improving the physical knowledge about the problem domain.

%%%%%%%%%%%%%%%%%%%%%%%%
\subsection{Classification statistics}
\label{sec:statistics}
In this work, the performance of the classification models is based on some statistical estimators, extracted from a matrix known as \emph{confusion matrix} \cite{Stehman1997}.

\begin{table}
\centering
\begin{tabular}{l|l|@{\hspace{0.7em}}c @{\hspace{1.7em}}cc}
\multicolumn{2}{c}{}& \multicolumn{2}{c}{\bfseries Predicted}&\\
\cline{3-4}
\multicolumn{2}{c}{} & \bfseries P=0 & \bfseries N=1 {\hspace{1.7em}}\\
\cline{3-4}
\multirow{2}{*}{\bfseries Target} {\hspace{0.3em}} & {\hspace{0.8em}} \bfseries p=0 {\hspace{0.8em}} & $a_{00}$ & $a_{10}$ {\hspace{0.7em}}\\
& {\hspace{0.8em}} \bfseries n=1 {\hspace{0.8em}} & $a_{01}$ & $a_{11}$ {\hspace{0.7em}}\\
\cline{3-4}
\multicolumn{1}{c}{} 
\end{tabular}
\caption{Example of a binary confusion matrix.}
\label{confmat}
\end{table}

The example shown in Table~\ref{confmat} is a confusion matrix for a binary classification. Each entry
$a_{ij}$ in this table is the number of records from class i predicted to be of class j. The numbers $a_{00}$ and $a_{11}$ show correct classified records. The $a_{01}$ records named \emph{False Positive} indicate wrong records classified in class 0, when their correct classification was class 1; instead, $a_{10}$ named \emph{False Negative} show the records classified in class 1 but belonging to class 0.
The total number of correct predictions is $a_{11} + a_{00}$, and the total number of wrong ones is $a_{10} + a_{01}$.
For a better comparison between different models, summarizing the results through a confusion matrix is the common way. We can do this using a \emph{performance metric}, such as \emph{accuracy}, defined as follows:

\begin{equation*}
Accuracy=\frac{a_{00}+a_{11}}{a_{00}+a_{11}+a_{01}+a_{10}}
\end{equation*}

A highest accuracy is the target of every classifier. Other important statistical estimators, for a better understanding of the results for each class, are:

\begin{equation*}
Purity=\frac{True Positive}{True Positive+ False Positive}
\end{equation*}

\begin{equation*}
Completeness=\frac{True Positive}{True Positive+ False Negative}
\end{equation*}

\begin{equation*}
Contamination=1-Purity=\frac{False Positive}{True Positive +False Positive}
\end{equation*}

\begin{equation*}
F1_{Score}=\frac{2}{(Purity)^{-1}+(Completeness)^{-1}}
\end{equation*}

\emph{Purity} of a class is the percentage of correctly classified objects in that class, divided by the total classified objects in that class. Also named as precision of a class.

\emph{Completeness} of a class is the percentage of the correctly classified objects in that class divided by the total amount of objects belonging to that class. Also named as recall of a class.

\emph{Contamination} of a class is the dual measure of purity.

\emph{F1-Score} of a class is the harmonic mean between purity and completeness of that class and it is a measure of the average trade-off between purity and completeness.

%%%%%%%%%%%%%%%%%%%%%%%%
\section{Experiments}
\label{sec:Experiments}

In order to pursue the main goal of the present work, related to a deep analysis of SNe in terms of their classification and characterization of the parameter space required to recognize their different types, we relied on the two simulation datasets, one in particular developed and specialized within the LSST project (see Sections \ref{sec:Data-SNPhotCC} and \ref{sec:Data-PLAsTiCC}). 
We preferred a statistical approach, by mapping the light curves into a set of statistical features. The classification with statistical data have been performed through the comparison of different types of classifiers, respectively, Nadam, RMSProp, Adadelta and Random Forest.

A data pre-processing phase was carried out on the PLAsTiCC dataset, based on a pruning on the flux and related error, in order to reduce the amount of negative fluxes present within data, which could affect the learning capability of the machine learning models. On the SNPhotCC dataset, both the errors in the flux and the quantity of negative fluxes were such that it was not deemed necessary to perform the pruning. The curves in the PLAsTiCC dataset were selected in successive steps so as to minimize the presence of negative fluxes, reaching, where possible, a subset of about 35,000 light curves per type. In the SNPhotCC dataset, on the other hand, all the given $5088$ SN-Ia curves were selected and the type II curves were reduced so as to balance the classes; the other types of SNe have been discarded, due to their negligible amount available.

The sequence of classification experiments followed an incremental complexity, starting from the most simple exercise on the PLAsTiCC dataset, i.e. the separation between periodic and non-periodic objects (\textit{P Vs NP}), expected to be well classified due to their very different features within any parameter space. In terms of initial minimization of negative fluxes, it was decided to apply the following replacement: for each class of objects, the observations related to the same day were grouped, by taking the least positive flux value. This value has been replaced to all the negative fluxes of that day.

As expected, the classifiers revealed a high capability to disentangle periodic from non-periodic objects. Therefore, in all further experiments we excluded periodic sources, by focusing the exclusive attention to variable objects, increasing the complexity of classification, by considering different sub-classes of transients and evaluating the performances of the selected machine learning classifiers.

The next step was, in fact, to recognize the SNe from all the other non-periodic objects available in the dataset (\textit{SNe Vs All}). But, preliminarly, we tested different methods for replacing the negative fluxes. For instance, in addition to the first mentioned method (e.g. minimum positive flux extracted from observations within the same day), a second method was chosen, in which negative fluxes were replaced by the constant number $0.001$, considered as the absolute minimum flux emitted by the sources. We tried also a third method, in which the negative fluxes were simply excluded from the input dataset, without any replacement. In theory, such third method was considered the worst case, since it would cause a drastic reduction of the light curve sample available. As we will show, the second method (the constant minimum flux value), obtained the best classification performances for all classifiers. Therefore, it was used as the reference for all further classification experiments. 

The subsequent experiments concerned some fine classifications of most interesting SNe types, starting from the classic case of \textit{SNIa Vs SNII} types, followed by a mix of SNIa Vs Superluminous SNe I (\textit{SNIa Vs SL-I}), concluding with the most complex case, based on the multi-class experiment, in which we tried to simultaneously classify all six different types of SNe (\textit{six-class SNe}).

\begin{figure}\centering
\includegraphics[scale=0.7]{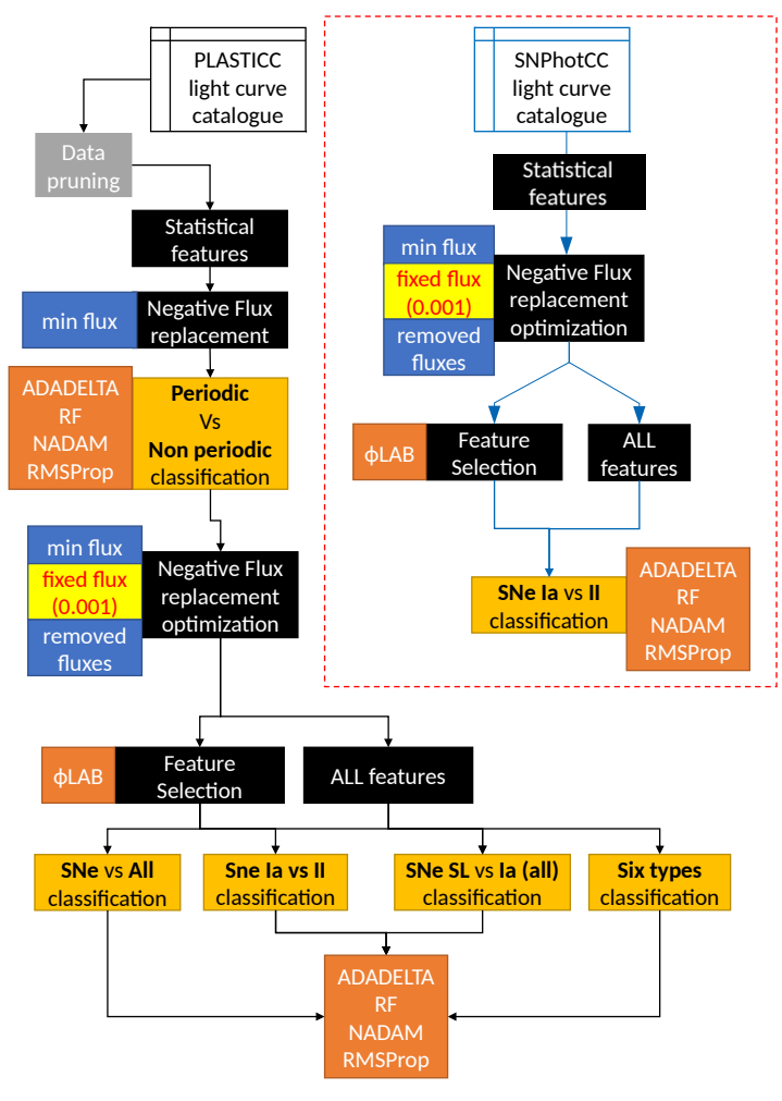}
\caption{Summary of the procedure designed and followed along the experiments.}
\label{sections}
\end{figure}

Besides the negative flux replacement, we investigated also the feature selection problem, in order to identify the most significant parameter space able to recognize different types of SNe. After the selection process we verified that such reduced amount of data dimensions could maintain sufficiently high the classification performances. 
We tried also to maintain uniform the number of features among the different use cases, although respecting their statistical importance, exploring the possibility to find a common parameter space, suitable for all classification cases. 

The \textit{SNIa Vs SNII} use case was also performed on the SNPhotCC dataset, since this dataset was composed almost exclusively by such two types of SNe. The results were then compared with those performed on the PLAsTiCC dataset, deprived of the \textit{u} and \textit{y} bands for uniformity with the SNPhotCC dataset bands, in order to maximize the fair comparison.

In summary, in this work five series of experiments were performed on the PLAsTiCC dataset and one on the SNPhotCC dataset. Such experiments were chosen hierarchically and considering the most important goal of this work, i.e. the fine classification of SNe types. An overview of the followed procedure is shown in Fig.~\ref{sections}.

%%%%%%%%%%%%%%%%%%%%%%%%
\subsection{Data pre-processing}
\label{preprocessing}
From the whole PLAsTiCC dataset a maximum of $200,000$ objects per class was randomly extracted (whenever possible). For each class, a pruning in flux and its error was performed. While, no any pruning was done on the SNPhotCC dataset. The Table~\ref{47} shows the limits derived from pruning.

\begin{table}\centering
	%\resizebox{\columnwidth}{!}{
\begin{tabular}{|*{8}c|}
\hline
Object & Band & Flux & Flux Error & Object & Band & Flux & Flux Error\\ 
\hline
%\endfirsthead
%\hline
%\endhead
%\hline \multicolumn{8}{c}{\textit{Continued on next page}} \\
%\endfoot
%\endlastfoot
\multirow{4}{*}{AGN} & u & $>$ -50 & $<$160 & \multirow{4}{*}{M-Dwarf}& u & $>$-60 & $<$300\\
 & g,r & $>$-50 & $<$160 & & g,r & $>$-60 & $<$100\\
 & i,z & $>$-50 & $<$160 & & i,z & $>$-60 & $<$80\\
 & y & $>$-50 & $<$160 & & y & $>$-60 & $<$180\\
\hline
\multirow{6}{*}{E. Binary} & u & $>$-200 & $<$800 & \multirow{6}{*}{Kilonova}& u & $>$-10 & $<$60\\
 & g & $>$-800 & $<$800 & & g & $>$-10 & $<$20\\
 & r & $>$-900 & $<$800 & & r & $>$-10 & $<$20\\
 & i & $>$-800 & $<$800 & & i & $>$-10 & $<$25\\
 & z & $>$-1100 & $<$800 & & z & $>$-20 & $<$40\\
 & y & $>$-800 & $<$650 & & y & $>$-30 & $<$70\\
\hline
\multirow{6}{*}{Mirae} & u & $>$-30 & $<$2500 & \multirow{6}{*}{$\mu$ Lens}& u & $>$-40 & $<$1700\\
 & g & $>$-20 & $<$800 & & g & $>$-20 & $<$250\\
 & r & $>$-50 & $<$900 & & r & $>$-30 & $<$400\\
 & i & $>$-1200 & $<$1700 & & i & $>$-40 & $<$300\\
 & z & $>$-8000 & $<$3000 & & z & $>$-60 & $<$400\\
 & y & $>$-11000 & $<$3300 & & y & $>$-90 & $<$500\\
\hline
\multirow{6}{*}{RR Lyrae} & u & $>$-1300 & $<$1500 & \multirow{6}{*}{SN Ia}& u & $>$-50 & $<$1350\\
 & g & $>$-6000 & $<$1500 & & g & $>$-20 & $<$500\\
 & r & $>$-6000 & $<$1500 & & r & $>$-20 & $<$400\\
 & i & $>$-4500 & $<$1500 & & i & $>$-40 & $<$170\\
 & z & $>$-4500 & $<$1200 & & z & $>$-60 & $<$200\\
 & y & $>$-5500 & $<$1200 & & y & $>$-100 & $<$300\\
\hline
\multirow{6}{*}{SN Iax} & u & $>$-30 & $<$550 & \multirow{6}{*}{SN Ia91bg}& u & $>$-30 & $<$800\\
 & g & $>$-10 & $<$150 & & g & $>$-20 & $<$200\\
 & r & $>$-20 & $<$150 & & r & $>$-20 & $<$200\\
 & i & $>$-30 & $<$100 & & i & $>$-30 & $<$150\\
 & z & $>$-50 & $<$125 & & z & $>$-40 & $<$150\\
 & y & $>$-90 & $<$200 & & y & $>$-90 & $<$325\\
\hline
\multirow{6}{*}{SN Ibc} & u & $>$-50 & $<$800 & \multirow{6}{*}{SN II}& u & $>$-40 & $<$200\\
 & g & $>$-20 & $<$200 & & g & $>$-20 & $<$100\\
 & r & $>$-20 & $<$150 & & r & $>$-20 & $<$100\\
 & i & $>$-30 & $<$100 & & i & $>$-30 & $<$100\\
 & z & $>$-60 & $<$125 & & z & $>$-60 & $<$100\\
 & y & $>$-110 & $<$350 & & y & $>$-110 & $<$150\\
\hline
\multirow{6}{*}{SL SN I} & u & $>$-30 & $<$1000 & \multirow{6}{*}{TDE}& u & $>$-20 & $<$200\\
 & g & $>$-10 & $<$150 & & g & $>$-10 & $<$50\\
 & r & $>$-15 & $<$125 & & r & $>$-10 & $<$50\\
 & i & $>$-20 & $<$100 & & i & $>$-20 & $<$50\\
 & z & $>$-40 & $<$100 & & z & $>$-30 & $<$75\\
 & y & $>$-70 & $<$175 & & y & $>$-60 & $<$150\\
\hline
\end{tabular}
\caption{Table of values retained after data pruning on the classes of PLAsTiCC dataset.}
\label{47}
\end{table}

After this first skimming, the amount of objects for the various classes was reduced to a maximum of about $35,000$ curves. The reduction for classes with more than 35K objects was driven by the choice of the curves with the least number of observations with negative fluxes and with at least $6$ observations per band. 

Due to the residual presence of negative fluxes, we started their handling by trying the following replacement method. By considering all the curves of a class, we checked all the observations of a given day. If in that day there was a negative or zero flux, then it was replaced with the lowest positive flux present. Else if only negative fluxes were present, they were replaced with the lowest positive flux of the previous day. This replacement has been applied to every day, for all curves and for all classes. An example of the replacing method is shown in Table~\ref{replacing}.

\begin{table}
  \centering
  %\tiny
 \begin{tabular}{|*{4}c|} 
 \hline
ID & MJD & \multicolumn{2}{c|}{Flux} \\ 
 & & Before & After\\
 \hline\hline
1 & $59820.0015$ & $-25.154862$ & $0.284215$ \\ 
 2 & $59820.0238$ & $15.458932$ & $15.458932$ \\ 
 3 & $59820.1234$ & $-5.848961$ & $0.284215$ \\ 
 4 & $59820.4451$ & $-20.548951$ & $0.284215$ \\ 
 5 & $59820.8251$ & $0.284215$ & $0.284215$ \\ 
 6 & $59820.0234$ & $-9.542318$ & $0.284215$ \\ 
 7 & $59820.6234$ & $10.854215$ & $10.854215$ \\ 
 \hline
\end{tabular}
\caption{Example of the negative fluxes replacement within the PLAsTiCC catalogue.}
\label{replacing}
\end{table}

Since $19$ features have been chosen for our statistical approach, by considering $6$ bands in PLAsTiCC, a total of $114$ features composed the original parameter space.

After the composition of statistical datasets, some light curves included some missing entries, or $NaN$ (Not-a-Number), causing the exclusion of those objects from the datasets, due to their unpredictable impact on the training of classifiers. The total amount of light curves per class is reported in Table~\ref{49}.

\begin{table}
  \centering
  %\tiny
 \begin{tabular}{|*{5}c|} 
 \hline
Dataset & Object & Curves & Object & Curves \\ 
 \hline\hline
\multirow{7}{*}{PLAsTiCC} & AGN & 34666 & E. Binary & 34484 \\ 
& Kilonova & 232 & M-Dwarf & 34849 \\ 
& Mirae & 1154 & $\mu$ Lens & 1187 \\ 
& RR Lyrae & 32698 & SN Ia & 34953 \\ 
& SN Iax & 34977 & SN Ia 91bg & 34923\\ 
& SN Ibc & 34932 & SN II & 34828 \\ 
& SL SN I & 34959 & TDE & 14023\\ 
& \multicolumn{2}{c}{\textbf{Total objects}} & \textbf{361711} &\\
\hline\hline
SNPhotCC & SNIa & 5088 & SNII & 12027\\
& \multicolumn{2}{c}{\textbf{Total objects}} & \textbf{17115} &\\
 \hline
\end{tabular}
\caption{Summary of the light curves composing the simulated datasets.}
\label{49}
\end{table}

%%%%%%%%%%%%%%%%%%%%%%%%
\subsection{Periodic Vs Non Periodic}
\label{periodicExp}
This was the first classification experiment, performed only on PLAsTiCC simulation. Having no need, at this level, to optimize the treatment of negative fluxes, we used only the method previously described (Sec.~\ref{preprocessing}). We had RR lyrae, Mirae variables and Eclipsing Binaries in the periodic class ($P$) and all the others in the non periodic ($NP$) class. To balance the classes we excluded some objects in the second class, as shown in Table~\ref{preprocesscurve}. The random partitioning percentage between training and test sets was fixed, respectively, to 80\% and 20\%.

\begin{table}
  \centering
  %\tiny
 \begin{tabular}{|*{6}c|} 
 \hline
\multirow{2}{*}{Object} & \multicolumn{2}{c}{Number of curves}& \multirow{2}{*}{Object} & \multicolumn{2}{c|}{Number of curves}\\ & Training & Test & & Training & Test\\
 \hline\hline
RR Lyrae & 26158 & 6540 & Kilonova & 187 & 46 \\ 
 E. Binary & 27587 & 6897 & M-Dwarf & 6001 & 1501\\ 
 Mirae & 923 & 231 & $\mu$ Lens & 950 & 238\\ 
 AGN & 6001 & 1501 & SN Ia & 6001 & 1501\\ 
 SN Iax & 6001 & 1501 & SN Ia 91bg & 6001 & 1501 \\ 
 SN Ibc & 6001 & 1501 & SN II & 6001 & 1501\\ 
 SL SN I & 6001 & 1501 & TDE & 6001 & 1501\\ 
 \hline\hline
 \multicolumn{2}{|c}{Total P Training} & 54668 & \multicolumn{2}{c}{Total NP Training} & 55146 \\
 \multicolumn{2}{|c}{Total P Test} & 13668 & \multicolumn{2}{c}{Total NP Test} & 13793 \\
 \hline
\end{tabular}
\caption{Summary of the sources belonging to the PLAsTiCC dataset in the $P$ (periodic class) Vs $NP$ (non periodic class) use case divided in training (80$\%$) and test (20$\%$) sets.}
\label{preprocesscurve}
\end{table}

\begin{table}
  \centering
  %\tiny
 \begin{tabular}{|*{6}c|} 
 \hline
\% & type & RF & Nadam & RMSProp & Adadelta\\
 \hline\hline
Accuracy                      & -  & 99 & 97 & 98 & 96\\ 
\multirow{2}{*}{Purity}       & NP & 99 & 97 & 99 & 95\\ 
                              & P  & 99 & 98 & 98 & 97\\ 
\multirow{2}{*}{Completeness} & NP & 99 & 98 & 98 & 97\\ 
                              & P  & 99 & 97 & 99 & 95\\ 
\multirow{2}{*}{F1 Score}     & NP & 99 & 98 & 98 & 96\\ 
                              & P  & 99 & 97 & 98 & 96\\ 
 \hline
\end{tabular}
\caption{Summary of the best results (in percentages) for the $4$ classifiers in the classification experiment \textit{P Vs NP}. For Nadam, RMSProp and Adadelta models, a decay value of $10^{-5}$ and a learning rate of $0.0005$ were assigned.}
\label{411}
\end{table}

This series of experiments, as expected, being the simplest given the intrinsic difference of the objects involved, did not reveal any surprise.
All estimators showed a great efficiency to recognize periodic objects from the variables (non periodic) ones.

%%%%%%%%%%%%%%%%%%%%%%%%
\subsection{Handling of negative fluxes}
\label{negFluxes}
In both simulated catalogues, as introduced in Sec.~\ref{sec:Experiments}, the presence of negative fluxes required an investigation on how to replace them in order to minimize their negative impact on the learning efficiency of machine learning models. Therefore, it was decided to approach this problem in three ways. 

The first (named as $M1$) was to replace their value as introduced in Sec.~\ref{sec:Experiments} (and preliminarly used for the Periodic Vs Non Periodic classification experiment, described in Sec.~\ref{periodicExp}): for each class of objects, the observations related to the same day were grouped, by taking the least positive flux value. This value has been replaced to all the negative fluxes of that day. 

The second approach (named as $M2$) was to replace the negative fluxes with a constant value of $0.001$, considered as the minimum flux emitted by the sources.

The third solution ($M3$) consisted into the total rejection of negative fluxes from the dataset, without any replacement. 

The impact on classification accuracy has been analyzed by comparing the three solutions in the \textit{SNe Vs All} (the class $All$ includes the rest of transient types) classification experiment on the Plasticc dataset and the \textit{SNIa Vs SNII} experiment on the SNPhotCC dataset. In both cases, the data have been treated with the three replacement types, producing different amount of objects per class. The entire composition of the datasets for the three methods is shown in Table~\ref{curvemetodi}, while the composition of the classes of SN, All, SNIa and SNII are shown in Table~\ref{curvemetodi2}.

\begin{table}
    \centering
  \resizebox{\textwidth}{!}{  
 \begin{tabular}{|*{9}c|} 
 \hline
& Object & \multicolumn{3}{c}{Number of curves} & Object & \multicolumn{3}{c|}{Number of curves} \\ 
& & $M1$ & $M2$ & $M3$ & & $M1$ & $M2$ & $M3$ \\
\hline\hline
\multirow{6}{*}{PLAsTiCC}&AGN & 34666 & 34666 & 34082 & Kilonova & 232 & 232 & 229\\ &$\mu$ Lens & 1187 & 1187 & 1144 & M-Dwarf & 34849 & 34849 & 34191\\ 
&SN Ia & 34953 & 34891 & 34423 & SL SN I & 34959 & 34959 & 34750\\ 
&SN Iax & 34977 & 34977 & 34680 & SN Ia 91bg & 34923 & 34923 & 34559\\ 
&SN Ibc & 34932 & 34932 & 34437 & SN II & 34828 & 34771 & 34393 \\ 
&TDE & 14023 & 14023 & 13985 & & & & \\ 

\multicolumn{2}{|c}{Total objects $M1$:}& 294529 & \multicolumn{2}{c}{Total objects $M2$:}& 294410 & \multicolumn{2}{c}{Total objects $M3$:}& 290873\\
\hline\hline

SNPhotCC & SNIa & 5088 & 5088 & 5086 & SNII & 5088 & 5088 & 5077\\ 
\multicolumn{2}{|c}{Total objects $M1$:}& 10176 & \multicolumn{2}{c}{Total objects $M2$:}& 10176 & \multicolumn{2}{c}{Total objects $M3$:}& 10163\\
 \hline
 \end{tabular}}
\caption{Summary of sources of datasets for each replacing method adopted for negative fluxes.}
\label{curvemetodi}
\end{table}

\begin{table}
    \centering
    %\tiny
 \begin{tabular}{|*{8}c|} 
 \hline
& & \multicolumn{2}{c}{$M1$} & \multicolumn{2}{c}{$M2$} & \multicolumn{2}{c|}{$M3$}\\
& Object & Training & Test & Training & Test & Training & Test\\
\hline\hline
\multirow{11}{*}{PLAsTiCC}& AGN & 27732 & 6934 & 27732 & 6934 & 27266 & 6816 \\
& Kilonova & 186 & 46 & 186 & 46 & 183 & 46\\
&$\mu$ Lens & 949 & 238 & 949 & 238 & 915 & 229\\ 
& M-Dwarf & 27879 & 6970 & 27879 & 6970 & 27353 & 6838\\
&SN Ia & 12001 & 3001 & 11975 & 2994 & 11802 & 2954 \\ 
& SL SN I & 12001 & 3001 & 12001 & 3001 & 11935 & 2979\\
&SN Iax & 12001 & 3001 & 12001 & 3001 & 11900 & 2976 \\ 
& SN Ia 91bg & 12001& 3001 & 12001 & 3001 & 11866 & 2975\\
&SN Ibc & 12001 & 3001 & 12001 & 3001 & 11828 & 2951 \\ 
& SN II & 12001 & 3001 & 11983 & 2992 & 11835 & 2970 \\ 
&TDE & 11218 & 2805 & 11218 & 2805 & 11188 & 2797 \\ 

& Total SN & 72006 & 18006 & 71962 & 17990 & 71166 & 17805 \\
& Total All & 67964 & 16993 & 67964 & 16993 & 66905 & 16726\\
\hline\hline

\multirow{2}{*}{SNPhotCC} & SNIa & 4071 & 1017 & 4071 & 1017 & 4062 & 1016\\
& SNII & 4071 & 1017 & 4071 & 1017 & 4070 & 1015\\

 \hline
 \end{tabular}
\caption{Summary of sources of training and test sets for each negative flux replacing method.}
\label{curvemetodi2}
\end{table}

The results of the two experiments are shown, respectively, in Tables \ref{confmet1} and \ref{confmet2}. The results indicated that, on average, in the case of the PLAsTiCC dataset, the second method ($M2$) obtained a better accuracy, with some exception in favor of $M3$. In the case of SNPhotCC dataset, on the other hand, $M2$ and $M3$ resulted more close in terms of classification efficiency. Therefore, since we were mostly interested to directly compare the classification performances between the two datasets, by considering also the drastic reduction of available sources using the $M3$ method, we definitely selected and applied the $M2$ to both datasets.

\begin{table}
    \centering
    %\tiny
 \begin{tabular}{|*{8}c|} 
\hline
Dataset & Use case & Algorithm & Class & Estimator & $M1$ & $M2$ & $M3$\\ 
 \hline\hline
\multirow{24}{*}{PLAsTiCC} & \multirow{24}{*}{\textit{SNe Vs All}} & \multirow{6}{*}{RF} & \multirow{3}{*}{SN} &                                Purity  & 86 & 91 & 85\\
                                    & & & &          Completeness  & 94 & 93 & 91\\
                                    & & & &          F1-score & 90 & 92 & 88\\\cline{4-8}
                & & & \multirow{3}{*}{All} &          Purity & 93 & 92 & 90\\
                                     & & & &          Completeness & 83 & 90 & 83\\
                                    & & & &          F1-score & 88 & 91 & 86\\\cline{3-8}
& & \multirow{6}{*}{Nadam} & \multirow{3}{*}{SN} &    Purity & 77 & 84 & 83\\
                                           & & & &    Completeness & 82 & 78 & 85\\
                                        & & & &      F1-score & 79 & 81 & 84\\\cline{4-8}
                    & & & \multirow{3}{*}{All} &      Purity & 80 & 78 & 84\\
                                         & & & &      Completeness & 73 & 85 & 82\\
                                        & & & &      F1-score & 76 & 81 & 83\\\cline{3-8}
& & \multirow{6}{*}{RMSProp} & \multirow{3}{*}{SN} &  Purity & 85 & 89 & 87\\
                                             & & & &  Completeness & 83 & 89 & 91\\
                                        & & & &      F1-score & 84 & 89 & 89\\\cline{4-8}
                    & & & \multirow{3}{*}{All} &      Purity & 83 & 88 & 90\\
                                         & & & &      Completeness & 85 & 89 & 86\\
                                        & & & &      F1-score & 84 & 88 & 88\\\cline{3-8}
& & \multirow{6}{*}{Adadelta} & \multirow{3}{*}{SN} & Purity & 80 & 85 & 85\\
                                              & & & & Completeness & 84 & 86 & 87\\
                                        & & & &      F1-score & 82 & 86 & 86\\\cline{4-8}
                    & & & \multirow{3}{*}{All} &      Purity & 82 & 85 & 86\\
                                         & & & &      Completeness & 78 & 84 & 84\\
                                        & & & &      F1-score & 80 & 85 & 85\\
\hline
\end{tabular}
\caption{Comparison among the three replacing methods for negative fluxes on the PLAsTiCC dataset in the classification case \textit{SNe Vs All}. For Nadam, RMSProp and Adadelta a learning rate of $0.001$ and a decay value of $10^{-5}$ were chosen. The statistics are expressed in percentages.}
\label{confmet1}
\end{table}

\begin{table}
    \centering
    %\tiny
 \begin{tabular}{|*{8}c|} 
 \hline
Dataset & Use case & Algorithm & Class & Estimator & M1 & M2 & M3\\ 
 \hline\hline
 \multirow{24}{*}{SNPhotCC} & \multirow{24}{*}{\textit{SNIa Vs SNII}} & \multirow{6}{*}{RF} & \multirow{3}{*}{SNIa} &                              Purity & 91 & 95 & 91\\
                                                & & & & Completeness & 94 & 97 & 93\\
                                               & & & & F1-score & 93 & 96 & 92\\\cline{4-8}
                          & & & \multirow{3}{*}{SNII} & Purity & 94 & 97 & 93\\
                                                & & & & Completeness & 91 & 95 & 91\\
                                               & & & & F1-score & 92 & 96 & 92\\\cline{3-8}
   & & \multirow{6}{*}{Nadam} & \multirow{3}{*}{SNIa} & Purity & 86 & 91 & 92\\
                                                & & & & Completeness & 92 & 92 & 94\\
                                               & & & & F1-score & 89 & 92 & 93\\\cline{4-8}
                          & & & \multirow{3}{*}{SNII} & Purity & 91 & 92 & 94\\
                                                & & & & Completeness & 86 & 91 & 92\\
                                               & & & & F1-score & 88 & 91 & 93\\\cline{3-8}
 & & \multirow{6}{*}{RMSProp} & \multirow{3}{*}{SNIa} & Purity & 91 & 92 & 93\\
                                                & & & & Completeness & 93 & 96 & 94\\
                                               & & & & F1-score & 92 & 94 & 94\\\cline{4-8}
                          & & & \multirow{3}{*}{SNII} & Purity & 93 & 96 & 94\\
                                                & & & & Completeness & 91 & 92 & 93\\
                                               & & & & F1-score & 92 & 94 & 94\\\cline{3-8}
& & \multirow{6}{*}{Adadelta} & \multirow{3}{*}{SNIa} & Purity & 89 & 86 & 92\\
                                                & & & & Completeness & 92 & 88 & 92\\
                                               & & & & F1-score & 91 & 87 & 92\\\cline{4-8}
                          & & & \multirow{3}{*}{SNII} & Purity & 92 & 88 & 92\\
                                                & & & & Completeness & 89 & 85 & 92\\
                                               & & & & F1-score & 90 & 87 & 92\\\cline{3-8}
 \hline
\end{tabular}
\caption{Comparison among the three replacing methods for negative fluxes on the SNPhotCC dataset in the classification case \textit{SNIa Vs SNII}. For Nadam, RMSProp and Adadelta a learning rate of $0.001$ and a decay value of $10^{-5}$ were chosen. The statistics are expressed in percentages.}
\label{confmet2}
\end{table}

%%%%%%%%%%%%%%%%%%%%%%%%%%%%%%%%
\subsection{Optimization of the Parameter Space for transients}
\label{ParameterSpace}
\begin{figure}\centering
\includegraphics[width=\textwidth]{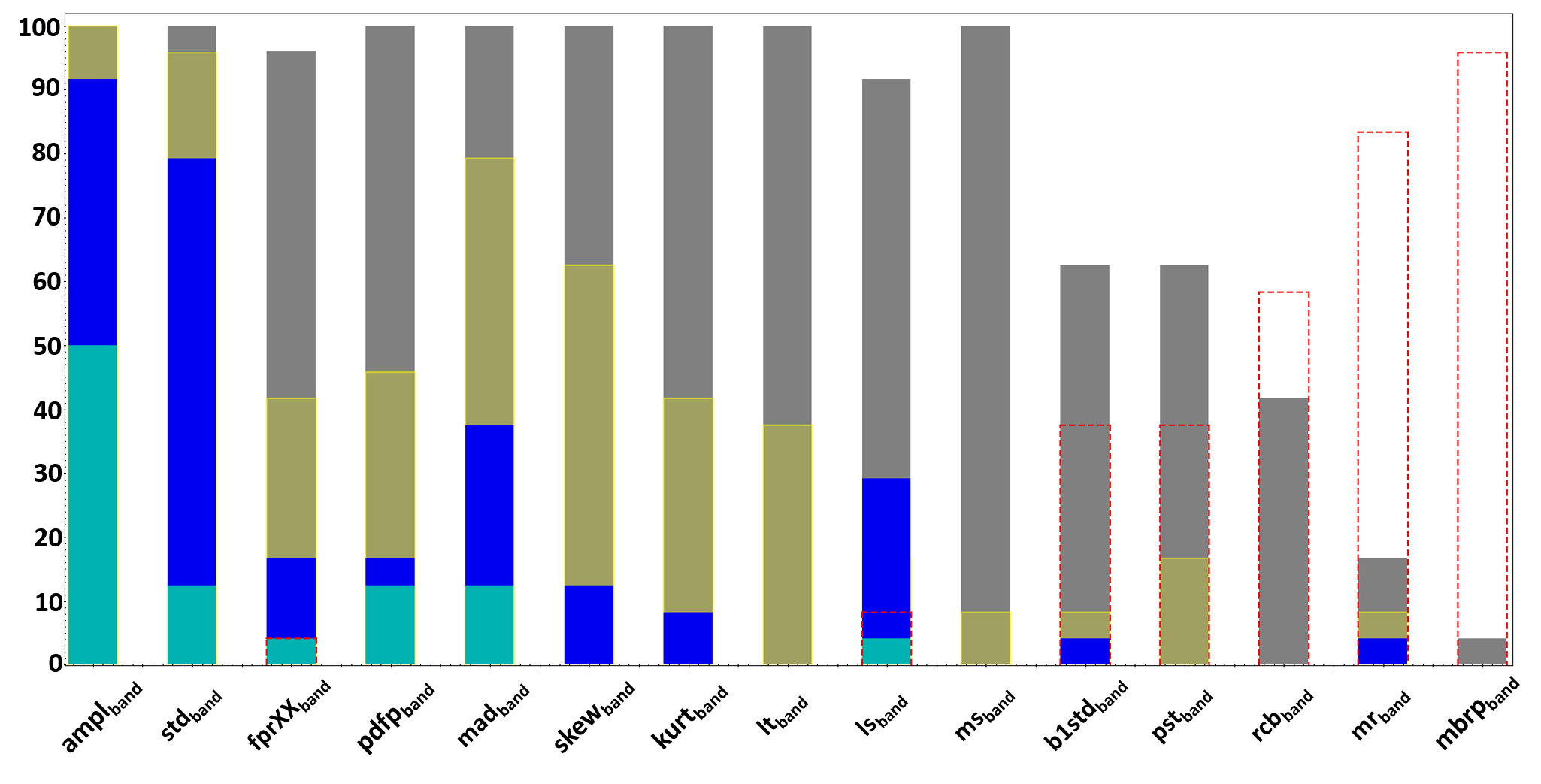}
\caption{Cumulative statistical analysis of the feature selection performed with the method $\Phi$LAB on different use cases. The results include the four classification cases on PLAsTiCC (\textit{SN Vs All}, \textit{SNIa Vs SNII}, \textit{SNIa Vs SL-I}, \textit{six-class SNe}) and the single case \textit{SNIa Vs SNII} on SNPhotCC. The indicated features are grouped per statistical type, including all their available bands. After having calculated the various feature rankings for each classification case with $\Phi$LAB, ordered by decreasing importance, the vertical bars shown in the histogram represent the percentage of common occurrences of each feature type, among various classification use cases, within, respectively, the first $25$\% (cyan), $50$\% (blue), $75$\% (yellow) and $100$\% (gray) of feature rankings. While dotted red bars indicate the percentage of common occurrences of rejection among various feature rankings.}
\label{fsresult}
\end{figure}

After choosing how to handle the negative fluxes, we investigated the statistical parameter space (PS) of the two simulated datasets, in order to explore the possibility to reduce the dimensionality of the classification problem (feature selection) and to analyze the impact of the resulting optimized PS on the classification efficiency for each particular type of classes involved in all cases, as well as the possibility to find a common set of relevant features, suitable to separate different types of transients. We applied the $\Phi$LAB algorithm, introduced in Sec.~\ref{sec:philab}, to both datasets in various classification use cases (except the preliminary experiment \textit{P Vs NP}), obtaining an optimized parameter space for each of them. The analysis of feature commonalities among all classification experiments is shown in Fig.~\ref{fsresult}. In particular, the feature selection of the \textit{SNIa Vs SNII} use case has been done on the PLAsTiCC dataset deprived of the \textit{u} and \textit{y} bands, for uniformity with the SNPhotCC dataset in terms of direct comparison.

%\begin{landscape}                                                           
%\end{landscape}

From the analysis of the histogram of Fig.~\ref{fsresult} it was possible to extract a common optimized parameter space, composed by relevant features with higher percentage of common occurrences among various classification use cases (the cumulative measurement process is explained in the caption of the Fig.~\ref{fsresult}). The extraction was done trying also to balance the different amount of relevant features provided by $\Phi$LAB in every classification case with their percentage of commonality among different cases, with the aim at extracting the same number of relevant features in all cases. The best compromise found is reported in Table~\ref{anfeat} and corresponds to $78$ extracted features (on a total of $114$) suitable for the six-band cases (\textit{ugrizy} in PLAsTiCC) and $52$ (on a total of $76$) for the four-band cases (\textit{griz} in SNPhotCC). These two resulting optimized (reduced) parameter spaces have been used in the classification cases described in the next sections, each time by comparing the classification efficiency between the complete and the reduced parameter spaces.  

\begin{table}
    \centering
    %\small
 \begin{tabular}{|*{7}c|} 
 \hline
Feature & [\textit{SNe Vs All}] & [\textit{SNIa Vs SNII}] & [\textit{SNIa Vs SL-I}] & [\textit{six-class SNe}] & | & [\textit{SNIa Vs SNII}] \\ 
        &   & & PLAsTiCC   & & | & SNPhotCC \\
 \hline\hline
ampl$_{band}$  & x & x & x & x & & x  \\ 
pdfp$_{band}$  & x & x & x &   & & x  \\
ms$_{band}$    &   &   &   & x & &   \\
mad$_{band}$   & x & x & x & x & & x  \\
std$_{band}$   & x & x & x & x & & x  \\
skew$_{band}$  & x & x & x & x & & x  \\
fprXX$_{band}$ & x & x & x & x & & x  \\
kurt$_{band}$  & x & x & x & x & & x  \\
ls$_{band}$    & x & x & x & x & & x  \\
lt$_{band}$    & x & x & x & x & & x  \\
 \hline
\textbf{Totals}  &   &  PLAsTiCC: \textbf{$78$} & &  SNPhotCC: \textbf{$52$} & & \\
 \hline
\end{tabular}
\caption{Summary of the resulting common optimized parameter spaces from the analysis of the feature selections. Each feature listed is intended to include all its available bands. First four use cases (columns $2$ to $5$) refer to the classification cases approached on PLAsTiCC with such optimized PS in six bands (\textit{ugrizy}), while last column is referred to the classification experiment done with SNPhotCC in four bands (\textit{griz}). For PLAsTiCC the optimized PS include $78$ features, while SNPhotCC is composed by $52$. Take into account that the feature fprXX$_{band}$ includes $5$ different types per band group (See Sec.~\ref{sec:Data-features} for details).}
\label{anfeat}
\end{table}

By looking at the optimized parameter spaces obtained (Fig.~\ref{fsresult}), extremely interesting is the presence of some common features among the various classification cases. In particular, the Amplitude (\textit{ampl}) shows a crucial role for the classification of various SNe types. Also important is the Standard Deviation (\textit{std}), which reaches $79.2\%$ of common occurrences. Equally interesting appears the high percentage of common rejections of Median Buffer Range Percentage (\textit{mbrp}), Magnitude Ratio (\textit{mr}) and R Cor Bor (\textit{rcb}). 
Within most of the light curves of the datasets used, the average value of the \textit{mbrp}, which is the percentage of points in an interval of $10\%$ of the median flux, is very high. This shows that most of the light curves are relatively contained within the flux extension.
The \textit{mr} feature, the percentage of points above the median magnitude, has always values greater than $40\%$, with a standard deviation of a lower order of magnitude, except in the case of the \textit{six-class SNe} problem, in which the standard deviation is comparable with the \textit{mr} value. This shows that most of the light curves are basically symmetrical in magnitude. Finally, the \textit{rcb} has an average value of about $30\%$ with a comparable standard deviation. Therefore, it ranges over the whole spectrum of possible values without any class distinction.

The \textit{ampl}, which from a physical point of view represents the half-amplitude, in magnitude, of the light curves, is the most important feature in all use cases and it is related to the different distribution of objects in the classes. In the \textit{SNe Vs All} use case, the class of SNe shows a bi-modal distribution, while the class \textit{All} shows an alternation between bi-modal and uni-modal distributions, with different peaks from the SNe distributions.
In the \textit{SNIa Vs SL-I} use case, the SNe Ia have a bi-modal distribution, unlike the SL-I type, which instead is uni-modal. The \textit{six-class SNe} use case shows that the SNe Ia have a similar peak w.r.t. the sub-types \textit{Iax}, \textit{Iabg91}, \textit{SL} and the \textit{Ibc}. The SNe II instead, show an unexpected shape similarity with the SNe Ia in the PLAsTiCC simulation, and this should explain a classification efficiency in the \textit{SNIa Vs SNII} case smaller than what obtained on the SNPhotCC data (see Sec.~\ref{SNeIavsII}).

The \textit{std}, the deviation from the mean flux, has the same trend of the \textit{ampl}, with bi-modal and uni-modal distributions and with peaks at different values. 

The \textit{fpr}, the flux percentage ratio, related to the sampling of the light curve assuming a relevance with the higher flux values, shows that, in the \textit{SNe Vs All} case and with the \textit{griz} bands, there are two distributions with distinguishable peaks. In the \textit{six-class SNe} case, the \textit{riz} bands, with the wider flux ratios, contribute to solve the envelope of the $6$ classes. In the \textit{SNIa Vs SL-I} case, the different distributions can be particularly identified in the \textit{rizy} bands, again in the broader flux ratios such as $50$, $65$ and $80$. Finally, in the \textit{SNIa Vs SNII} problem the distinction is more complex and only in few \textit{riz} band cases it is possible to see the two different distributions.

In the other relevant features shown in Fig.~\ref{fsresult}, we do not infer distinct distributions in the various use cases, but only different fluctuations around the same distribution. This means that all the curves of all the classes share, more or less, the same distribution w.r.t. the flatness of the curve (\textit{kurt}), the symmetry of the curve (\textit{skew}), the slope deriving from the linear fit (\textit{lt}), the period obtained from the peak frequency of the Lomb Scargle Periodogram (\textit{ls}), the ratio of difference between percentiles and the median (\textit{pdfp}) and finally the median of deviations from the median (\textit{mad}). Since these features have proved to be highly relevant, this implies that those fluctuations in the class distributions contribute substantially to the classification of different types of SNe.
Finally, in the \textit{six-class SNe} problem, another feature appears relevant, which is the maximum difference in magnitude between two successive epochs (\textit{ms}), providing, slightly in the \textit{u} band and in a more consistent way in the \textit{y} one, fluctuations suitable in principle for the resolution of the more complex classification.

%%%%%%%%%%%%%%%%%%%%%%
\subsection{Supernovae versus All}
\label{SNevsAll}
In this use case we had SNe type Ia, Iax, Ia 91bg-like, Ibc, II and SL-I within the \textit{SNe} class and all the other object types, except the excluded periodic ones, in the \textit{All} class. We performed the experiments on the PLAsTiCC dataset with the $4$ classifiers using, respectively, the entire set of statistical features available ($114$) and with the optimized parameter space ($78$). The amount of objects for each type included in the two classes is shown in Table~\ref{417}. 

\begin{table}
    \centering
    %\tiny
 \begin{tabular}{|*{3}c|} 
 \hline
Type & Training & Test \\
 \hline\hline
SN Ia & 11975 & 2994 \\ 
 SN Iax & 12001 & 3001\\ 
 SN Ia91bg & 12001 & 3001\\ 
 SN Ibc & 12001 & 3001 \\
 SN II & 11983 & 2992\\ 
 SL SN I & 12001 & 3001\\ 
 Kilonova & 186 & 46\\ 
 M-Dwarf & 27879 & 6970\\ 
 $\mu$ Lens & 949 & 238\\ 
 TDE & 11218 & 2805 \\
 AGN & 27732 & 6934 \\ 
  \hline\hline
  Total SN & 71962 & 17990 \\
  Total All & 67964 & 16993 \\
 \hline
\end{tabular}
\caption{Summary of the objects belonging to the PLAsTiCC dataset, used for the \textit{SNe Vs All} experiment, randomly partitioned in training (80$\%$) and test (20$\%$) sets.}
\label{417}
\end{table}                            
                            
Among Nadam, RMSProp and Adadelta, the best performances were obtained with the RMSProp in both cases (whole and optimized parameter spaces). While Random Forest reached the best classification performances. The statistical results are shown in Table~\ref{418}.

\begin{table}
    \centering
    %\tiny
 \begin{tabular}{|*{10}c|} 
 \hline
 & & \multicolumn{2}{c}{Random Forest} & \multicolumn{2}{c}{Nadam} & \multicolumn{2}{c}{RMSProp} & \multicolumn{2}{c|}{Adadelta}\\ 
Features &                                   & All  & 78   & All  & 78   & All  & 78   & All  & 78\\ 
 \hline\hline
\% Accuracy & -                        & 92 & 92 & 85 & 86 & 90 & 90 & 86 & 85 \\ 
\multirow{2}{*}{\% Purity}       & SN  & 91 & 91 & 85 & 86 & 91 & 91 & 86 & 84 \\
                              & All & 92 & 92 & 85 & 86 & 89 & 90 & 86 & 86 \\
\multirow{2}{*}{\% Completeness} & SN  & 93 & 93 & 86 & 87 & 90 & 90 & 87 & 87 \\
                             & All  & 90 & 90 & 84 & 85 & 90 & 90 & 85 & 83 \\
\multirow{2}{*}{\% F1 Score}    & SN   & 92 & 92 & 86 & 87 & 90 & 87 & 87 & 86 \\
                             & All  & 91 & 91 & 84 & 86 & 90 & 86 & 86 & 84 \\ 
 \hline
\end{tabular}
\caption{Summary of the statistical results for the $4$ classifiers with, respectively, all the features and the $78$ selected. For Nadam, RMSProp and Adadelta, the values of $10^{-5}$ and $0.0005$ were assigned to the decay and learning rate hyper-parameters, respectively.}
\label{418}
\end{table}

%%%%%%%%%%%%%%%%%%%%%%
\subsection{Supernovae Ia versus II}
\label{SNeIavsII}
In this experiment we considered only SNe of type Ia and II. In this case it was possible to use both SNPhotCC and PLAsTiCC datasets, since in the case of SNPhotCC, these two types of SN were available.
The amount of objects used is shown in Table~\ref{425}.  

\begin{table}
    \centering
    %\tiny
 \begin{tabular}{|*{4}c|} 
 \hline
\multirow{2}{*}{Dataset} & \multirow{2}{*}{Type} & \multicolumn{2}{c|}{Number of curves}\\  & & Training & Test \\
 \hline\hline
\multirow{2}{*}{PLAsTiCC} & SN Ia & 27964 & 6990 \\ 
& SN II & 27983 & 6966 \\ 
& Total & 55947 & 13956 \\
 \hline\hline
\multirow{2}{*}{SNPhotCC} & SN Ia & 4071 & 1017 \\ 
& SN II & 4071 & 1017 \\
& Total & 8142 & 2034 \\
 \hline
\end{tabular}
\caption{Summary of the objects belonging to the datasets used for the \textit{SNIa Vs SNII} experiment on PLAsTiCC and SNPhotCC, randomly partitioned in training (80$\%$) and test (20$\%$) sets.}
\label{425}
\end{table}

We performed the experiment with the $4$ classifiers using, respectively, all the features and the amounts related to the two optimized feature sets, respectively, $78$ for PLAsTiCC and $52$ for SNPhotCC. For a direct comparison between the SNPhotCC and PLAsTiCC datasets, we also considered a reduced version of the PLAsTiCC dataset, by excluding the \textit{u} and \textit{y} bands for uniformity with the SNPhotCC catalogue in terms of bands available. The statistical results are reported in Table~\ref{426}.

\begin{table}
    \centering
    %\tiny
 \begin{tabular}{|*{22}c|}
 \hline
 & & \multicolumn{5}{c}{Random Forest} & \multicolumn{5}{c}{Nadam} & \multicolumn{5}{c}{RMSProp} & \multicolumn{5}{c|}{Adadelta}\\
                              &      & \multicolumn{3}{c}{PLA} & \multicolumn{2}{c}{SNP} & \multicolumn{3}{c}{PLA} & \multicolumn{2}{c}{SNP} & \multicolumn{3}{c}{PLA} & \multicolumn{2}{c}{SNP} & \multicolumn{3}{c}{PLA} & \multicolumn{2}{c|}{SNP}\\
Bands used                    &      & 6   & 6  & 4  & 4   & 4  & 6   & 6  & 4  & 4   & 4  & 6   & 6  & 4  & 4   & 4  & 6   & 6  & 4  & 4   & 4\\
Features                      &     & All & 78 & 52 & All & 52 & All & 78 & 52 & All & 52 & All & 78 & 52 & All & 52 & All & 78 & 52 & All & 52\\ 
 \hline\hline
\% Accuracy                   & -    & 78  & 79 & 78 &  96 & 96 & 71  & 72 & 71 & 93  & 94 & 76  & 76 & 78 & 94  & 96 & 74  & 74 & 73 & 90  & 95\\ 
\multirow{2}{*}{\% Purity}    & Ia   & 76  & 76 & 76 &  95 & 95 & 70  & 70 & 69 & 90  & 92 & 74  & 74 & 75 & 93  & 94 & 72  & 73 & 72 & 89  & 93\\ 
                              & II   & 81  & 81 & 80 &  97 & 97 & 72  & 73 & 74 & 95  & 96 & 78  & 78 & 80 & 95  & 98 & 75  & 74 & 75 & 92  & 96\\ 
\multirow{2}{*}{\% Completeness} & Ia   & 82  & 83 & 82 &  97 & 97 & 75  & 76 & 77 & 96  & 96 & 79  & 80 & 82 & 95  & 98 & 76  & 75 & 77 & 92  & 96\\ 
                              & II   & 74  & 74 & 74 &  95 & 95 & 67  & 67 & 65 & 89  & 92 & 73  & 71 & 73 & 93  & 93 & 71  & 72 & 70 & 88  & 93\\ 
\multirow{2}{*}{\% F1 Score}  & Ia   & 79  & 79 & 79 &  96 & 96 & 72  & 73 & 73 & 93  & 94 & 77  & 77 & 79 & 94  & 96 & 74  & 74 & 74 & 90  & 95\\ 
                              & II   & 77  & 78 & 77 &  96 & 96 & 70  & 70 & 69 & 92  & 94 & 75  & 74 & 76 & 94  & 95 & 73  & 73 & 72 & 90  & 95\\ 
 \hline
\end{tabular}
\caption{Summary of the statistical results for the $4$ classifiers in the \textit{SNIa Vs SNII} experiment. For each classifier it is reported the statistics related to the PLAsTiCC (PLA columns) and SNPhotCC (SNP columns) datasets. In the case of PLAsTiCC, the columns are related to the whole original feature space (All) and the optimized one ($78$) using 6 bands (\textit{ugrizy}), together with the reduced feature space ($52$) using $4$ bands (\textit{griz}) for a direct comparison with the corresponding optimized parameter space obtained on SNPhotCC. For Nadam, RMSProp and Adadelta, the values of $10^{-5}$ and $0.0005$ were assigned to, respectively, the decay and learning rate hyper-parameters, in the cases of  $78$ features. While $10^{-5}$ and $0.001$ values have been assigned for the cases with $52$ features.}
\label{426}
\end{table}               

In terms of classification performance, it appears evident the discrepancy between the two datasets. The capability of classifiers to recognize the two classes is higher on SNPhotCC and this implies a strong dependency of learning models from the overall accuracy of the simulations. Furthermore, the very similar percentages among the whole feature set and the optimized versions probes the capability of the feature selection method $\Phi$LAB to extract a set of relevant features, able to preserve the level of classification efficiency.   
                                                                             
%%%%%%%%%%%%%%%%%%%%%%
\subsection{Superluminous SNe versus SNe I}
\label{SLvsSNI}
In the \textit{SNIa Vs SL-I} experiment, the three sub-classes of SNe, Ia, Ia91bg and Iax have been mixed in the same percentage and then classified against Superluminous SNe I. We performed the experiments with the $4$ classifiers using all the features and the $78$ selected with $\Phi$LAB. The amount of objects per type is shown in Table~\ref{456}. 

\begin{table}
    \centering
    %\tiny
 \begin{tabular}{|*{3}c|} 
 \hline
\multirow{2}{*}{Type} & \multicolumn{2}{c|}{Number of curves}\\  & Training & Test \\
 \hline\hline
SN Ia & 9323 & 2331 \\ 
  SN Iax & 9323 & 2331 \\ 
  SN Ia91bg & 9323 & 2331 \\ 
  SLSN I & 27967 & 6992 \\ 
  \hline\hline
  Total Ia & 27969 & 6993 \\
  Total SL & 27967 & 6992 \\
 \hline
\end{tabular}
\caption{Summary of the objects belonging to the dataset used for the \textit{SNIa Vs SL-I} experiment on PLAsTiCC, randomly partitioned in training (80$\%$) and test (20$\%$) sets.}
\label{456}
\end{table}

The statistical results of the classification are shown in Table~\ref{457}.

\begin{table}
    \centering
    %\tiny
 \begin{tabular}{|*{10}c|} 
 \hline
 & & \multicolumn{2}{c}{Random Forest} & \multicolumn{2}{c}{Nadam} & \multicolumn{2}{c}{RMSProp} & \multicolumn{2}{c|}{Adadelta}\\ 
Features &                               & All & 78 & All & 78 & All & 78 & All & 78\\ 
 \hline\hline
\% Accuracy                      & -     & 88  & 87 & 81  & 82 & 85  & 82 & 71  & 70 \\ 
\multirow{2}{*}{\% Purity}       & SL-I  & 83  & 80 & 77  & 74 & 81  & 77 & 71  & 70 \\
                                 & SN Ia  & 93  & 93 & 85  & 89 & 90  & 87 & 71  & 70 \\
\multirow{2}{*}{\% Completeness} & SL-I  & 94  & 95 & 87  & 92 & 91  & 89 & 71  & 70 \\
                                 & SN Ia  & 80  & 76 & 74  & 69 & 79  & 74 & 72  & 71 \\
\multirow{2}{*}{\% F1 Score}     & SL-I  & 88  & 87 & 82  & 82 & 86  & 83 & 71  & 70 \\
                                 & SN Ia  & 86  & 84 & 79  & 78 & 84  & 80 & 71  & 70 \\ 
 \hline
\end{tabular}
\caption{Summary of the statistical results for the $4$ classifiers on the \textit{SNIa Vs SL-I} experiment, with, respectively, all the features and the $78$ of the optimized parameter space of PLAsTiCC dataset. For Nadam, RMSProp and Adadelta, the values of $10^{-5}$ and $0.0005$ were assigned to the decay and learning rate hyper-parameters, respectively.}
\label{457}
\end{table}

By analyzing the results, it is noticeable the lower performance of Adadelta w.r.t. other classifiers, where Random Forest appeared the best one for all estimators. The similar results obtained for both parameter spaces confirm the validity of the feature selection.

In terms of error percentages on the SN I class (all Ia sub-types), Table~\ref{4424} reports the level of contamination for each sub-type in the experiment with all features and using Random Forest. 

\begin{table}
    \centering
    %\tiny
 \begin{tabular}{|*{5}c|} 
 \hline
\multirow{2}{*}{Class} & \multirow{2}{*}{Total} & Correctly  & \multicolumn{2}{c|}{\multirow{2}{*}{Wrongly classified}} \\
      &       & classified & & \\
      \hline\hline
 SN Ia & 2331 & 2328 & 3 & $\approx$ 0\% \\ 
 SN Iax & 2331 & 1508 & 823 & 35\% \\ 
 SN Ia91bg & 2331 & 1779 & 552 & 24\% \\ 
 \hline
\end{tabular}
\caption{Summary of the contamination analysis among all SN Ia sub-types obtained by the Random Forest, with the complete parameter space, on the \textit{SNIa Vs SL-I} experiment.}
\label{4424}
\end{table}

As shown in Table~\ref{4424}, the most contaminated sub-class is SNIax, which indicates its high difficulty of recognition among other SN types.

%%%%%%%%%%%%%%%%%%%%%%
\subsection{Simultaneous classification of six SNe sub-types}
\label{sixClasses}
Last classification experiment performed was the most complex, because we tried to classify simultaneously all the six classes of SNe available in the PLAsTiCC dataset. The experiments with the $4$ models were performed using all the features and the $78$ selected by the optimization procedure. The amount of objects per class is shown in Table~\ref{463}.

\begin{table}
    \centering
    %\tiny
 \begin{tabular}{|*{3}c|} 
 \hline
\multirow{2}{*}{SN Class} & \multicolumn{2}{c|}{Number of curves}\\  & Training & Test \\
 \hline\hline
 Ia     & 27912 & 6979 \\ 
 Ia91bg & 27938 & 6985 \\ 
 Iax    & 27981 & 6996 \\ 
 II     & 27816 & 6955 \\ 
 Ibc    & 27945 & 6987 \\ 
 SL I   & 27967 & 6992 \\ 
 \hline
\end{tabular}
\caption{Summary of the objects belonging to the dataset used for the \textit{six-class SNe} experiment on PLAsTiCC, randomly partitioned in training (80$\%$) and test (20$\%$) sets.}
\label{463}
\end{table}

The statistical results of the six-class classification is reported in Table~\ref{464}.

\begin{table}
    \centering
    %\tiny
 \begin{tabular}{|*{10}c|} 
 \hline
  & & \multicolumn{2}{c}{Random Forest} & \multicolumn{2}{c}{Nadam} & \multicolumn{2}{c}{RMSProp} & \multicolumn{2}{c|}{Adadelta}\\
Features                         &             & All & 78 & All & 78 & All & 78 & All & 78 \\ 
 \hline\hline
\% Accuracy                      & -           & 66  & 62 & 53  & 55 & 60  & 61 & 48  & 48\\ 
\hline
\multirow{6}{*}{\% Purity}       & SN Ia       & 79  & 79 & 68  & 71 & 73  & 76 & 62  & 59 \\ 
                                 & SN Ia 91bg  & 82  & 78 & 64  & 70 & 79  & 81 & 52  & 58 \\
                                 & SN Iax      & 58  & 57 & 46  & 48 & 52  & 51 & 39  & 34 \\
                                 & SN II       & 74  & 75 & 58  & 61 & 68  & 66 & 56  & 55 \\
                                 & SN Ibc      & 40  & 42 & 32  & 34 & 34  & 35 & 32  & 32 \\
                                 & SL SN I     & 62  & 59 & 48  & 47 & 56  & 56 & 47  & 50 \\
                                 \hline
\multirow{6}{*}{\% Completeness} & SN Ia       & 77  & 77 & 56  & 57 & 68  & 67 & 55  & 48 \\ 
                                 & SN Ia 91bg  & 25  & 30 & 27  & 20 & 20  & 17 & 21  & 16 \\
                                 & SN Iax      & 33  & 37 & 16  & 20 & 26  & 27 & 21  & 15 \\
                                 & SN II       & 79  & 79 & 79  & 77 & 77  & 78 & 67  & 67 \\
                                 & SN Ibc      & 64  & 57 & 47  & 47 & 54  & 58 & 38  & 47 \\
                                 & SL SN I     & 91  & 91 & 76  & 88 & 88  & 85 & 85  & 87 \\
                                 \hline
    \multirow{6}{*}{\% F1 Score} & SN Ia       & 78  & 78 & 62  & 63 & 71  & 71 & 58  & 53 \\ 
                                 & SN Ia 91bg  & 39  & 44 & 38  & 31 & 32  & 28 & 30  & 26 \\
                                 & SN Iax      & 42  & 45 & 23  & 29 & 35  & 35 & 27  & 21 \\
                                 & SN II       & 76  & 77 & 67  & 68 & 72  & 72 & 61  & 60 \\
                                 & SN Ibc      & 49  & 48 & 38  & 40 & 42  & 44 & 34  & 38 \\
                                 & SL SN I     & 73  & 72 & 59  & 62 & 68  & 67 & 61  & 63 \\
 \hline
\end{tabular}
\caption{Summary of the statistical results for the $4$ classifiers on the \textit{six-class SNe} experiment, with, respectively, all the features and the $78$ of the optimized parameter space of PLAsTiCC dataset. For Nadam, RMSProp and Adadelta, the values of $10^{-4}$ and $0.001$ were assigned to the decay and learning rate hyper-parameters, respectively.}
\label{464}
\end{table}

Also in this case, the Random Forest obtained best results and the similar statistics between the whole and optimized parameter space confirm the good performances of the feature selection method.
By analyzing the classification estimators for the single classes, the SNIa91bg showed a high difficulty to be recognized, while Ia91bg and Iax types were often confused for SNIbc. SL type resulted the most complete, although the purity was reduced by the contamination of SNIbc and SNIax (Tab.~\ref{cont}). 

%\begin{figure}
%    \centering
%    \includegraphics[scale=0.45]{figures/contamination}
%    \caption{Percentages of contamination in the \textit{six-class SNe} classification results.}
%    \label{contold}
%\end{figure}

\begin{table}
    \centering
    %\tiny
 \begin{tabular}{cc|*{6}c|c|c} 
 
   \multicolumn{9}{c}{\textbf{Random Forest Classification}}& \parbox[t]{2mm}{\multirow{12}{*}{\rotatebox[origin=c]{90}{\red{Contamination \%}}}} \\
   \multicolumn{9}{c}{Predicted \%} \\
   \multicolumn{2}{c}{}&\textbf{Ia}&\textbf{Iabg}&\textbf{Iax}&\textbf{II}&\textbf{Ibc}&\multicolumn{1}{c}{\textbf{SL}}\\\cline{3-9}
\multirow{6}{*}{\shortstack{T\\r\\u\\e}}&\textbf{Ia}&\textbf{77.3}&-&-&22.6&-&0.1&\red{22.7}\\
&\textbf{Iabg}&-&\textbf{25.3}&17.5&-&47.6&9.6&\red{74.7}\\
&\textbf{Iax}&-&2.5&\textbf{32.6}&0.01&45.1&19.7&\red{67.4}\\

&\textbf{II}&21.0&-&-&\textbf{78.7}&0.01&0.3&\red{21.3}\\
&\textbf{Ibc}&0.1&3.0&6.0&0.03&\textbf{63.9}&27.0&\red{36.1}\\
&\textbf{SL}&0.1&0.1&0.3&4.6&4.1&\textbf{90.8}&\red{9.2}\\\cline{3-9}
\end{tabular}
    \caption{Percentages of contamination in the \textit{six-class SNe} classification results.}
    \label{cont}
\end{table}

Finally SNIa and SNII types, although reducing their efficiency w.r.t. the dedicated two-class experiment, maintained a sufficient level of classification.

%%%%%%%%%%%%%%%%%%%%%%
\section{Discussion and conclusions}
\label{discussion}
The present work is related to the important problem of classification of astrophysical variable sources, with special emphasis to SNe. Their relevance in terms of cosmological implications is well known, causing a special attention to the problem of recognizing different types of such  astronomical explosive events.

To face this challenge, the SNPhotCC dataset and the PLAsTiCC dataset have been chosen to have a statistical sample, albeit of simulations, as wide as possible. Based on the objects in the datasets, a test campaign with increasing complexity has drawn up. To approach the problem we have chosen $4$ machine learning methods that require a transformation of light curves into a series of statistical features, potentially suitable to recognize different source types.

In the construction of statistical datasets, the presence of negative fluxes within the observations had to be solved, due to their negative impact on the learning capability of ML models.
Working directly with the light curves, their shape is relevant, thus the presence of negative fluxes is not a big problem, because it is always possible to translate the curve along the ordinate axis. In the statistical parameter space instead, since there are features requiring the conversion to magnitudes and since the translation would alter the features values in an unpredictable way, the negative fluxes must be replaced in some way. To solve this problem we tried  three approaches, as described in Sec.~\ref{negFluxes}. In the first one, the atmospheric and instrumental setup conditions were respected, by grouping the observations taken in the same day; this solution evidently introduced noise, by altering the phase within groups of light curves. The second solution, which proved to be the best candidate, replaced negative fluxes with a positive number, by maintaining  unchanged the sampling, and introducing a lower contribution of noise within data. Finally, the third method removed the observations with negative fluxes, thus highly sub-sampling the light curves. From the classification results obtained adopting the second solution, we were confident that the deformations undergone by the light curves were not able to alter their original nature nor to significantly reduce the performances in both simulation datasets used.

The parameter space analysis was approached with the $\Phi$LAB algorithm to perform a reduction of dimensionality of the classification use cases and to investigate the possibility to identify a common set of features that could be considered suitable to recognize different types of transients. From the comparison between the original and optimized feature spaces, in terms of classification performance, the adopted method resulted extremely reliable to find a reduced set of relevant features, able to preserve the amount of information required to maintain the same level of classification efficiency. Starting from the $\Phi$LAB results, a statistical analysis was performed, which highlighted some interesting aspects related to the physical nature of transients and SNe in particular. The Amplitude feature, representing the semi-difference between the minimum and maximum of the light curve, resulted the most relevant. Since the various classes of SNe have different light peaks, the semi-difference of the amplitude of the curve is typical of each different type of object. Also relevant resulted the standard deviation, MAD, and all features related to the percentiles or characterizing the light curve shape, such as skewness and kurtosis. The relevance of percentiles is related to the different decay time of the light radiation for the various types of SNe. Although a SN is not a periodic event, the feature related to the Lomb-Scargle periodogram has a high importance, because it is able to classify the SNe with a different periods of light decay. On the other hand, all the feature related with thresholds on the number of points around the median (such as the rcb, mr and mbrp), were rejected by our feature analysis method, probably due to their average values too close to their limits.

The most important outcome of the parameter space analysis was the identification of a feature set common to all classification use cases that revealed a coherent behavior in terms of classification performances obtained in all cases, always well close to the efficiency arising from the original parameter spaces.

In terms of pure classification among different types of sources, the high capability to distinguish between Periodic and Non Periodic objects confirmed what expected and posed ML methodology as a good candidate to approach the transient classification problem in Astronomy. 

\begin{figure}
    \centering
    \includegraphics[width=\textwidth]{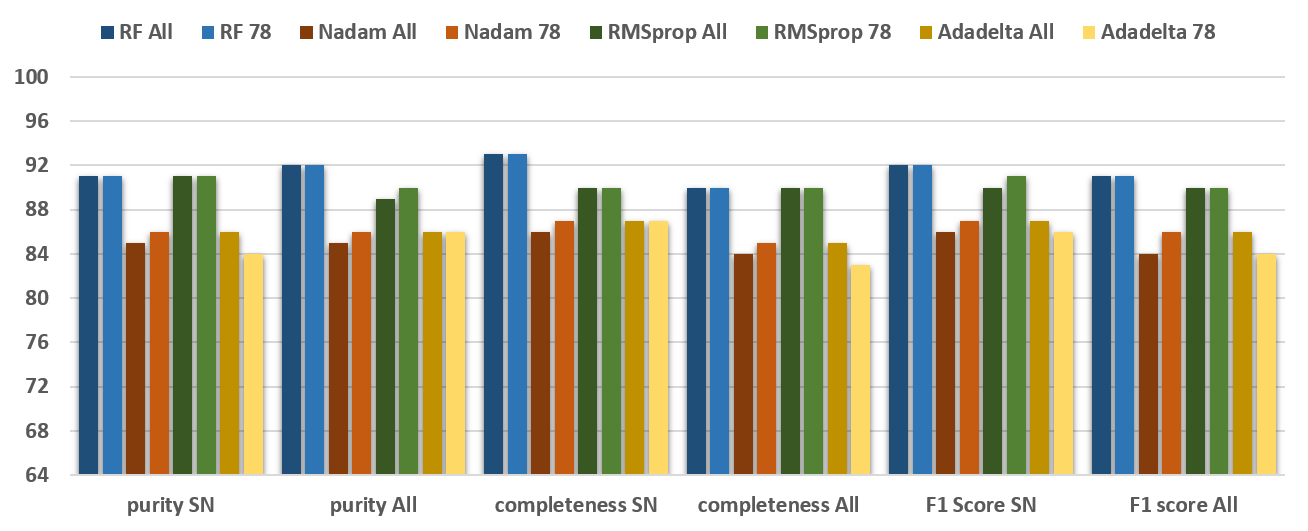}
    \caption{Histogram of the statistical results (in \%) for the \textit{SNe Vs All} classification problem.}
    \label{an2}
\end{figure}

Once removed periodic objects, the high completeness ($93\%$) reached in classifying SNe in the \textit{SNe Vs All} case (Fig.~\ref{an2}), confirmed that ML methods, in particular the Random Forest, could be suitable to distinguish SNe from other transients. 

We wanted to verify in the remaining $7\%$, which was the most contaminating among the different sub-types of SNe; both SNe Ia91bg and SNe Iax were found to have the highest misclassification rate ($12\%$ of their test set). Moreover, from this analysis it was revealed that SNIa and SNII have an error rate of about $1$ per thousand, a remarkable result compared to the other SNe error rates. For completeness, the contamination was also verified for the \textit{All} class, revealing that the AGN type has an error rate of $1$ per thousand, while the M-Dwarf and the TDE are the classes with the highest error rates ($16\%$). Further experiments should be carried out to identify the SNe classes with which these two different types of transients are confused and to verify which features play a key role in their classification. 

\begin{figure}
    \centering
    \includegraphics[width=\textwidth]{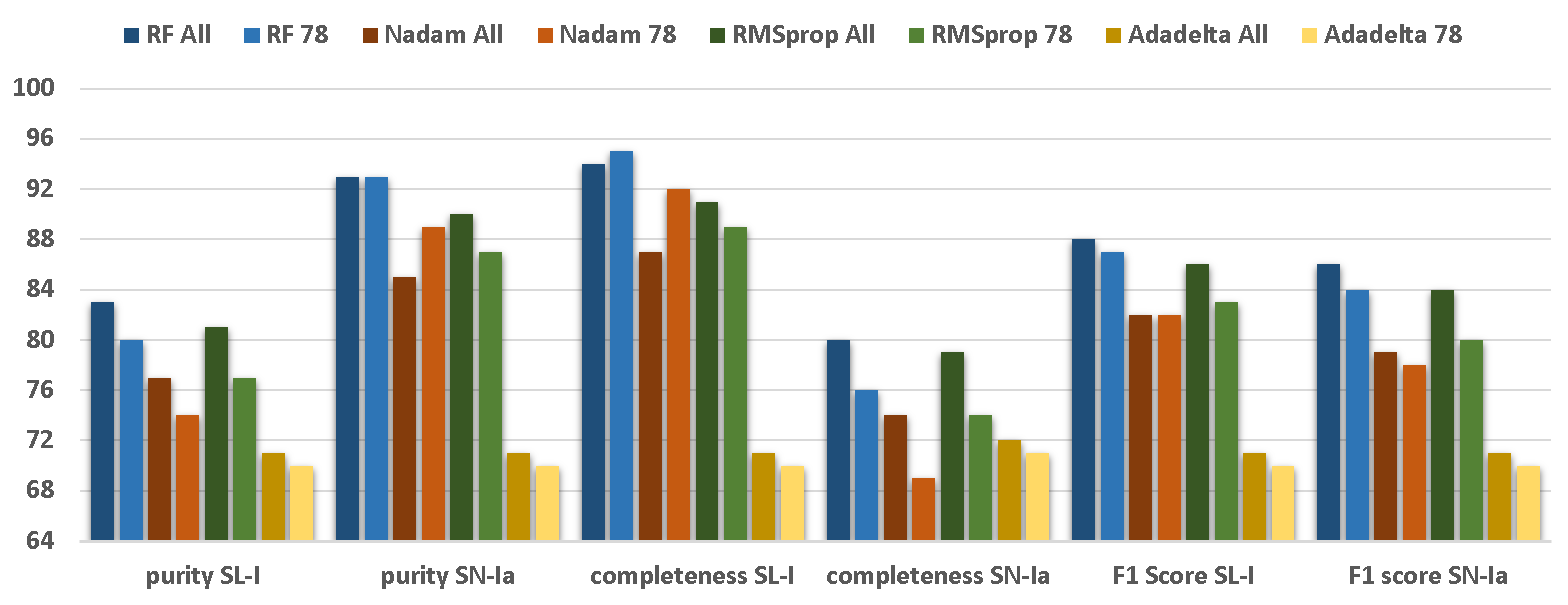}
    \caption{Histogram of the statistical results (in \%) for the \textit{SNIa Vs SL-I} classification problem.}
    \label{an6}
\end{figure}

Another interesting case was the classification between Super Luminous (SL) SNe and the mixed Ia types, from which ML appeared able to recognize the SL category with a completeness, in the best case, of $95\%$ (Fig.~\ref{an6}). 

Although further experiments could be in principle performed, from the results obtained in this work,  we can suppose to have identified a set of features suitable to help the classification of the SNe Ia, II and SL. However, those able to classify the sub-types Ia91bg and Iax are still unclear. Hopefully, with the availability of real LSST data in the near future, tests with only these two sub-classes could be conducted, with at most the addition of SNIbc, to evaluate which features could result relevant to recognize such types of SNe.

%%%%%%%%%%%%%%%%%%%%%%%%%%%%%%%%%%%%%%%%%%%%%%%%%%%%%%%%%%%%%%%%%%%%%%%%%%
\begin{acknowledgement}
The software package of machine learning models used in this work was developed within the DAME project \cite{Brescia2014}.
MB and GR acknowledge the financial contribution from the agreement \textit{ASI/INAF 2018-23-HH.0, Euclid ESA mission - Phase D}, while MB acknowledges also the
\textit{INAF PRIN-SKA 2017 program 1.05.01.88.04} and the \textit{MIUR Premiale 2016: MITIC}.
Topcat \cite{Taylor05} has been used for this work.
\end{acknowledgement}

\bibliographystyle{spphys}
\bibliography{main}

\end{document}